\documentclass[12pt,preprint]{aastex}
\newcommand{\boldlambda}{\mbox{\boldmath$\lambda$}}
\newcommand{\boldphi}{\mbox{\boldmath$\phi$}}
\newcommand{\boldr}{\mbox{\boldmath$r$}}

\begin{document}

\title{
NONLINEAR EVOLUTION OF GLOBAL HYDRODYNAMIC SHALLOW-WATER INSTABILITY
IN SOLAR TACHOCLINE
}
\author{MAUSUMI DIKPATI}
\affil{High Altitude Observatory, National Center for Atmospheric
Research \footnote{The National Center
for Atmospheric Research is sponsored by the
National Science Foundation. },
3080 Center Green, Boulder, Colorado 80301; dikpati@ucar.edu}

\begin{abstract}

We present a fully nonlinear hydrodynamic 'shallow water' model of the 
solar tachocline. 
The model consists of a global spherical shell of differentially
rotating fluid, which has a deformable top, thus allowing motions 
in radial directions along with latitude and longitude directions. 
When the system is perturbed, in the course of its nonlinear evolution
it can generate unstable low-frequency shallow-water shear 
modes from the differential rotation, high-frequency gravity waves, 
and their interactions. Radiative and overshoot tachoclines are
characterized in this model by high and low effective gravity values 
respectively. Building a semi-implicit spectral scheme containing 
very low numerical diffusion, we perform nonlinear evolution of 
shallow-water modes. Our first results show that, (i) high latitude
jets or polar spin-up occurs due to nonlinear evolution of unstable 
hydrodynamic shallow-water disturbances and differential rotation, 
(ii) Reynolds stresses in the disturbances together with changing 
shell-thickness and meridional flow are responsible for the evolution
of differential rotation, (iii) disturbance energy primarily remains 
concentrated in the lowest longitudinal wavenumbers, (iv) an oscillation 
in energy between perturbed and unperturbed states occurs due to 
evolution of these modes in a nearly dissipation-free system, and 
(v) disturbances are geostrophic, but occasional nonadjustment in 
geostrophic balance can occur, particularly in the case of high effective
gravity, leading to generation of gravity waves. We also find that a
linearly stable differential rotation profile remains nonlinearly stable.

\end{abstract}

\section{Introduction}

The solar tachocline plays a fundamental role in governing the dynamics 
and evolution of the Sun's interior magnetic fields, 
manifestations of which are observed at the solar surface. The tachocline
is a very thin ($\sim 0.03R$) transition layer at the base of the 
convection zone through which the solar differential rotation changes 
from latitudinal in the convection zone to a uniform 
core rotation. This layer contains strong radial and weak
latitudinal differential rotation. 

Due to existence of both radial and latitudinal shear in the tachocline,
the study of its dynamics and physics is complex. Analysis of tachocline 
hydro- and magnetohydrodynamics have proceeded in two different ways:
the first way treats the thin tachocline explicitly as a full three dimensional
multilayered spherical system (see, for example, \citet{clz2008}). The other
way is to treat the tachocline as a one-layer system. The multilayer system 
has been used to study the dynamics of both shears, radial and latitudinal, 
while the one-layer system has been applied to examine the dynamics and 
stability of only the latitudinal shear. Over the past decade, various 
models have been built for global HD/MHD of the tachocline latitudinal 
shear -- 2D, quasi-3D shallow-water systems and 3D thin-shell primitive 
equation systems, under both the Boussinesq and hydrostatic approximations. 

The linear and nonlinear calculations of HD/MHD instability of tachocline 
latitudinal shear have been performed in great detail for a 2D 
(latitude-longitude) system. These studies led to the following major 
outcomes: (i) solar tachocline latitudinal shear is hydrodynamically 
stable in pure 2D systems \citep{watson81,dk87,cdg99,garaud01}; (ii) 
MHD shear instability with longitudinal wavenumber $m=1$ to 7 sets in
and persists in the solar tachocline in the presence of a wide range 
of toroidal field profiles (from broad to narrow bands), amplitudes and 
latitude-locations \citep{gf97,dg99,gd02}; (iii) axisymmetric HD/MHD instability
does not, or more precisely, cannot exist in 2D; (iv) as consequences of the 
nonlinear evolution of this instability, broad toroidal fields open
up into "clam-shell"  patterns \citep{cally01}, whereas narrow bands tip 
\citep{cdg03}.

We now know that 
the 2D results are robust. All of them obtained from linear analysis
have been found to appear in more complex quasi-3D `shallow-water' systems
\citep{dg01,gd02,dgr03}, in which the third dimension is generally 
included in a restricted way, namely the upper boundary of the shallow 
fluid-shell is allowed to vary with latitude, longitude and time so that 
the radial motions can occur. Intuitively it is expected that the 
inclusion of the third dimension would produce some other unstable
modes that are purely characteristic of 3D because, in addition to 
the kinetic and magnetic energy reservoirs due to the 
presence of differential rotation and magnetic field, 3D systems allow 
energy extraction from the potential energy reservoir, particularly when
the effective gravity ($G$) of the system is low. $G$ is a measure of 
the subadiabatic stratification of the fluid layer; low $G$ can allow
more deformation of the fluid-layer's top surface. In linear studies, 
it was indeed found that a hydrodynamically stable 2D tachocline can be 
unstable to quasi-3D perturbations, due to the presence of this additional 
energy source in a shallow-water model \citep{dg01}.

Many idealized shallow-water models have been built over the past century,
since the first, single-layer model by Hough (1898). For studying the 
mid-latitude oceanic tides and atmospheric zonal flows and meridional
circulation, multi-layer shallow-water models are used \citep{hp98,
hh08}. Shallow water models are also being used for operational forecasts 
of El Nino and La Nina \citep{pf03}. 

Shallow water theory was applied to the global solar tachocline  
by Gilman (2000), who first formulated an MHD analog of a 
shallow-water system. Since 
then, all global studies of HD/MHD shallow-water instabilities in the solar 
tachocline have been performed in the linear regime. Nonetheless, even 
within the scope of a linear model, the upward bulging of a magnetized 
shallow fluid layer created at certain longitudes, by growing modes with 
low-longitudinal wavenumbers, can provide a plausible explanation of the 
Sun's ``active longitudes'' (see \citet{dg05} for details). 
This linear model was able to simulate active longitudes 
up to 15 Carrington Rotations, after which nonlinear interactions among the
unstable modes start playing important roles. 
In order to be able to simulate active longitudes for one full solar cycle
and several solar cycles, it is necessary to study the
the nonlinear evolution of the aforementioned instability in the
tachocline fluid layer.

Over the past decade, several 3D thin-shell and full 3D HD and MHD models 
for linear analysis of tachocline shear instabilities have been built 
\citep{cally01,zls03,asr05,gdm07}. However, unlike many studies of nonlinear
evolution of shallow-water instabilities in the oceanic and atmospheric
context (see, e.g. \citet{pf03b}), there exist only a few studies of the 
nonlinear evolution of 3D MHD instabilities in the solar context, using
3D thin-shell models \citep{mgd07,hc09}. Although in a shallow-water model 
the radial dimension of the fluid layer is included in a simplified way 
by allowing the layer's thickness to vary with latitude, longitude and 
time, it is not as restricted 
compared to a full 3D MHD system as it might appear. \citet{rd03} have shown
that an MHD shallow-water approach is a first-order approximation of a full
MHD equilibrium in which the magnetic curvature stress is balanced by a
latitudinal pressure gradient for a strongly subadiabatic stratification
and by modification in the shear flow for a weakly subadiabatic stratification. 
The height deformation of the top surface layer in the shallow-water approach 
corresponds to the deformation of the surfaces of constant entropy required 
for latitudinal force balance in the full MHD approach. Thus using a
shallow water system for the Sun, specifically for the solar tachocline,
can help reduce the complexity of a full 3D MHD system
without much affecting the basic physics behind the global dynamics of the
fluid layer.

Putting aside the added complexity due to the presence of magnetic fields,
we first focus on 
the nonlinear evolution of shear instability in an unmagnetized solar
tachocline and find out how the tachocline latitudinal differential 
rotation profile can evolve with time and get modified due to the 
influence of this instability. In pure 2D calculations, the jets were
seen to form around the locations of the singular points arising from the
coincidence between the Doppler-shifted phase-speed of the unstable modes 
and the shear-speed of the system, or the Alfvenic singular points in the
case of magnetized tachocline \citep{cdg04}. Along with
the exploration of jet-formation on unperturbed differential rotation profiles
due to the nonlinear evolution of hydrodynamic shear instability, another 
major aim of this paper is to study the nonlinear evolution of the bulging
and depression of the fluid layer in conjunction with the shear flow pattern,
or in other words, the exchanges between the kinetic and potential energies
of the system. 

We will present the mathematical formulation including equations and 
solution techniques in \S2 and in the Appendix, and the implementation 
of spectral viscosity 
in \S3. In the description of results in \S4 we will present in the first 
subsection how the energy spectrum develops, i.e. how the total energy gets
distributed in different longitudinal wavenumbers and how they evolve 
with time. The subsequent subsections of \S4 will focus on the evolution
of the Sun's longitude-averaged differential rotation and the height
profile, the energy exchanges between kinetic and potential energy 
reservoirs, and the evolution of flow and deformation of the fluid layer on
the latitude-longitude surface. We will summarize our findings in \S5.

\section{Mathematical Formulation}

\subsection{Governing equations}
 
Hydrodynamic shallow-water equations were built for studying global atmospheric
and oceanic dynamics; they exist in the literature in various forms, for example 
see \citet{pedlosky87}. Following that, \citet{dg01} described the hydrodynamic 
shallow-water equations, along with the usual assumptions and approximations in 
their \S2.1 and \S2.2 and solved them in the linear regime. Briefly, in a 
shallow water model the velocity (${\bf V}$) of the fluid can be defined as 
${\bf V}= u{\hat{ \boldlambda}} + v {\hat{ \boldphi}} + w{\hat{ \boldr}}$,
where $u$ and $v$ are the horizontal velocity components in longitude
($\lambda$) and latitude ($\phi$) respectively, and $w$ is the radial velocity. 
In the shallow water approximation, $u$ and $v$ are independent of depth, and
$w$ is a linear function of depth. For a shallow fluid layer with a rigid 
bottom and a deformable top, the maximum vertical velocity will be at the top.

Due to the thinness of the tachocline compared to the solar radius, the 
divergence of radius and the density variation within the shallow fluid layer 
are ignored in the momentum and mass-continuity equations. All time scales of 
the system are considered much longer than the acoustic time scale. With these
assumptions, the hydrodynamic shallow water equations for a single-layer 
tachocline can be derived from the mass continuity and momentum equations
(see \citet{dg01} for detailed derivation). In order to study the nonlinear 
evolution of the hydrodynamic shallow-water instability in the solar
tachocline, we start from the equations (4), (5) and (8) of \citet{dg01}
and rewrite the nonlinear dimensional hydrodynamic shallow-water equations 
in the rotating frame as,

$${\partial u \over \partial t} = {v \over r_0 \cos\phi}
\left[{\partial v \over \partial\lambda} - {\partial \over
\partial\phi} (u \cos\phi)\right] - {1 \over r_0 \cos\phi} {\partial
\over \partial\lambda}\left( {u^2 + v^2 \over 2}\right)
-{gH \over r_0 \cos\phi} {\partial h \over \partial\lambda}
+2 \omega_c v \sin\phi, \quad\eqno(1) $$

$${\partial v \over \partial t} = -{u \over r_0 \cos\phi}
\left[{\partial u \over \partial\lambda} - {\partial \over
\partial\phi} (u \cos\phi)\right] - {1 \over r_0 } {\partial
\over \partial\phi}\left( {u^2 + v^2 \over 2}\right)
-{gH \over r_0 } {\partial h \over \partial\phi} 
- 2 \omega_c u \sin\phi, \quad\eqno(2) $$

$${\partial \over \partial t}(1+h) = -{1 \over r_0 \cos\phi}
{\partial \over \partial\lambda}\left((1+h)u\right) - {1 \over r_0 \cos\phi}
{\partial \over \partial\phi}\left((1+h)v \cos\phi\right),
\quad\eqno(3) $$
in which $u$ and $v$ are the horizontal velocity components in longitude
($\lambda$) and latitude ($\phi$), and $h$ is the deformable top surface.

We make equations (1-3) dimensionless by choosing the radius of the 
shell ($r_0$) as the length-scale and the inverse of the core rotation 
rate (${\omega_c}^{-1}$) as the time-scale of the system; thus the thickness
of the fluid layer representing either the radiative or the overshoot part
of the solar tachocline should be between 0.01 and 0.05 dimensionless
units, and one year time will correspond to about 100 dimensionless
units. The nondimensional equations are

$${\partial u \over \partial t} = {v \over \cos\phi}
\left[{\partial v \over \partial\lambda} - {\partial \over
\partial\phi} (u \cos\phi)\right] - {1 \over \cos\phi} {\partial
\over \partial\lambda}\left( {u^2 + v^2 \over 2}\right)
-G {1 \over \cos\phi} {\partial h \over \partial\lambda}
+2 \omega_c v \sin\phi, \quad\eqno(4) $$

$${\partial v \over \partial t} = -{u \over \cos\phi}
\left[{\partial u \over \partial\lambda} - {\partial \over
\partial\phi} (u \cos\phi)\right] - {\partial
\over \partial\phi}\left( {u^2 + v^2 \over 2}\right)
-G {\partial h \over \partial\phi}
- 2 \omega_c u \sin\phi, \quad\eqno(5) $$

$${\partial \over \partial t}(1+h) = -{1 \over \cos\phi}
{\partial \over \partial\lambda}\left((1+h)u\right) - {1 \over \cos\phi}
{\partial \over \partial\phi}\left((1+h)v \cos\phi\right),
\quad\eqno(6) $$
in which $G=gH/r_0^2\omega_c^2$ is the effective gravity. With typical values
$H$, the pressure scale height, the value of $G$ is high ($\ge 10$) in the 
radiative tachocline and low ($< 1$) in the overshoot part of the tachocline.

Due to the pole problem, it is difficult to solve the advective form 
of shallow-water equations in spherical polar coordinates. It is a fundamental 
problem for solving numerically the partial differential equations on the 
sphere, particularly using finite difference methods -- certain terms are 
likely to be unbounded in the neighborhood of the poles. Spectral methods
are popular for their high accuracy and their ability to avoid the pole 
problem. There exists a vast literature on how to avoid the pole problem. 
We follow here a technique proposed by \citet{swarzt96} and so we define the 
vorticity ($\zeta$) and divergence ($\delta$) as 
$$\zeta = {1 \over \cos\phi}\left[{\partial v \over \partial\lambda}-
{\partial \over \partial\phi}\left(u\cos\phi\right)\right], 
\quad\eqno(7a)$$
$$\delta = {1 \over \cos\phi}\left[{\partial u \over \partial\lambda}+
{\partial \over \partial\phi}\left(v\cos\phi\right)\right], 
\quad\eqno(7b)$$
and rewrite the equations (4-6) as follows:
$$ {\partial u \over \partial t} = \left(\zeta+f\right)v-{1 \over \cos\phi}
{\partial \over \partial\lambda}\left[gh+{1 \over 2}\left(u^2+v^2\right)
\right], \quad\eqno(8) $$
$$
{\partial v \over \partial t} = -\left(\zeta+f\right)u-
{\partial \over \partial\phi}\left[gh+{1 \over 2}\left(u^2+v^2\right)
\right], \quad\eqno(9)$$
$${\partial h \over \partial t} = -\left(1+h\right)\delta-
{u \over \cos\phi} {\partial h \over \partial\lambda} -
v {\partial h \over \partial\phi}, \quad\eqno(10)$$ 
where,$f = 2\omega_c\sin{\phi}$ is the Coriolis parameter.

We apply the spectral transform method \citep{swarzt96} to solve the Equations 
(8-10), by expanding the vector field (i.e. the flow) and the scalar field
(height) respectively in terms of vector and scalar harmonics. Thus the pole
problem can be handled efficiently without raising the order of the equations.
The appendix describes in detail the technique used to solve the Equations
(8-10).

\subsection{Description of Initial State, Numerical Algorithm and Spectral 
Viscosity}

We express the solar tachocline latitudinal differential rotation in angular
measure, $\omega$, so that $u=\omega\,\cos\phi$. From the knowledge of
helioseismic observations the mathematical form of $\omega$ has been adopted
as
$$\omega = s_0 - s_2 \mu^2 - s_4 \mu^4 - \omega_c, \eqno(11)$$
\noindent in which $\omega_c$ represents the solid rotation of the core, and 
$s_0$, $s_2$ and $s_4$ are numerical coefficients. $(s_2 + s_4)/s_0 \sim 0.3$ 
represents solar latitudinal differential rotation near the surface, whereas 
in the radiative part of the tachocline, $ (s_2 + s_4)/s_0 \le 0.17$ 
\citep{cdg99}. Detailed parameter
surveys in the linear regime \citep{dg01} have revealed that $\omega$ 
becomes unstable to perturbations with longitudinal wavenumbers $m=1$ and
$m=2$ for a wide range of $s_2$ and $s_4$ values. In order to focus our
study on how this instability evolves in the cases of high and low effective
gravity ($G$), we pick an $\omega$ with $s_2=0.15$, $s_4$=0.09 and $s_0=1.051$ 
so that the pole-to-equator differential rotation is 24\%. Linear studies
indicate that this differential rotation is hydrodynamically unstable for 
a wide range of $G$. Therefore we can demonstrate the features of the time-evolution
of this differential rotation for both high $G$ and low $G$ cases, in particular
$G=10$ and 0.5 \citep{rd03, dgr03} which are, respectively, characteristic
of radiative and overshoot tachoclines.

We also study a few other cases with 30\% and 15\% pole-to-equator differential 
rotation. In all cases we start with an initial reference state profile that
is a combination of unperturbed profile and a perturbation with low longitudinal
wavenumbers (mostly $m=1$ mode), with a specified amplitude with respect to 
the unperturbed differential rotation. 

In all simulations we start from known solutions to the linearized
instability equations, rather than from perturbations of arbitrary form,
which has the effect of initially filtering gravity waves out of the 
calculations. But once nonlinearities become important, gravity wave modes
can and frequently are excited and subsequently participate in the overall
dynamics.

\subsection{Spectral Viscosity}

Spectral methods, despite their accuracy and absence of pole problems,
are prone to generating spurious oscillations in physical space
and hence a reduction in global accuracy. In the case of nonlinear 
evolution equations in an idealized, inviscid model like the one we are
presently dealing with, spectral methods generate energy
cascade from lower to higher wave number modes, which often creates
a spectrum with an erroneously high amplitude tail at high longitudinal
wavenumbers -- the so-called spectral blocking problem, which could
eventually lead to a poor energy-conservation during the nonlinear evolution
of the system. The traditional approach is to use an artificial 
horizontal diffusion to deal with these difficulties of spectral 
methods.
 
\begin{figure}[hbt]
\epsscale{1.0}
\plotone{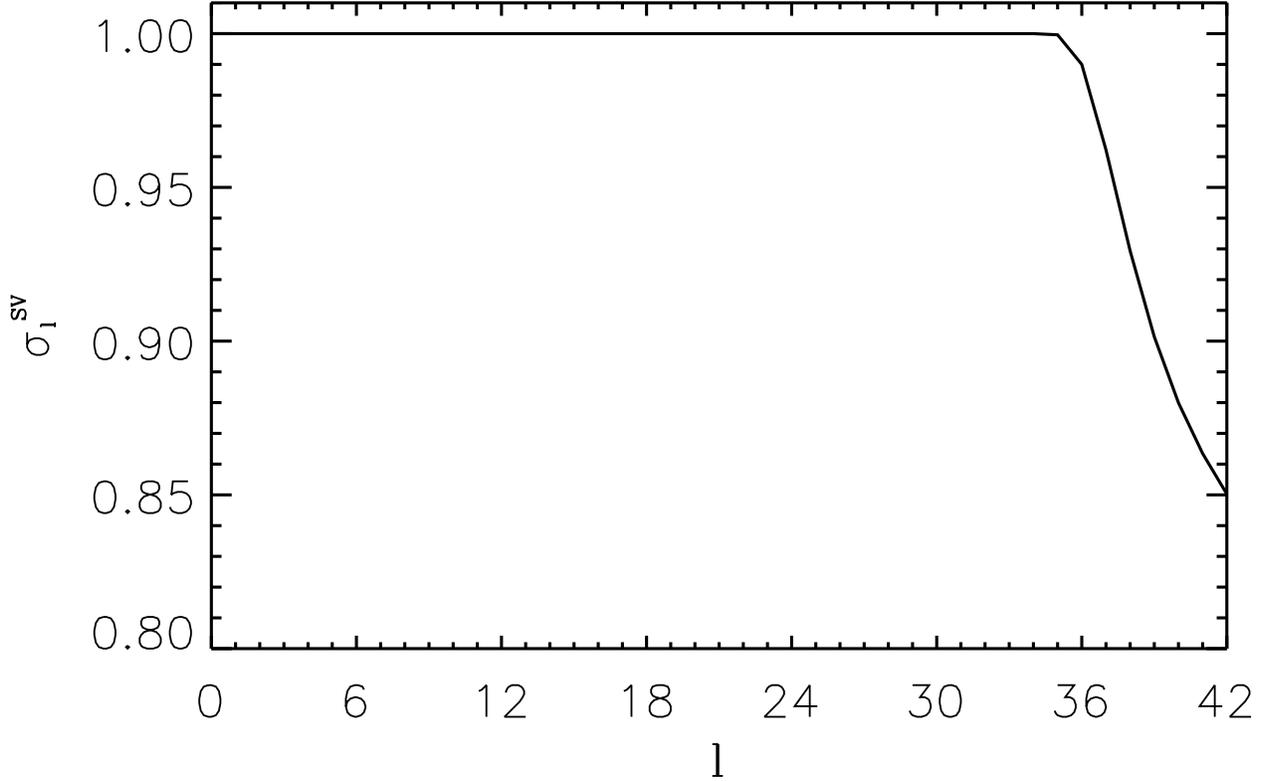}
\caption{Spherical spectral viscosity operator as function
of wavenumber $l$ in Legendre polynomial ${P_l}^m$.
}
\label{spectral-viscosity}
\end{figure} 

We apply the Spectral Viscosity (SV), as prescribed by \citet{gelb01},
which is a spherical spectral viscosity operator, able to conserve 
invariants of the system at no additional computing cost and to retain
the high wave number information in the system. The dynamical variables 
$a_r^{lm}$, $a_i^{lm}$, $b_r^{lm}$, $b_i^{lm}$, $c_r^{lm}$ 
and $c_i^{lm}$ are filtered at  each time step using the formula
$$a_r^{lm}=\sigma_l^{SV}a_r^{lm}, \quad\eqno(11a)$$
in which
$$\sigma_l^{SV}={1 \over 1+2\Delta_t\epsilon \hat{q_l}^2 l^2{(l+1)}^2},
\quad\eqno(11b)$$
with $\hat{q_l}$ satisfying the following two conditions:
$$\hat{q_l}= 0,\,\,\, {\rm for} \,\,  l\le l_c,$$
$$ \,\,\,\, =exp{\left[-\frac{(l-M)^2}{2{(l-l_c)}^2}\right]}, \,\,\, 
              {\rm for} \,\, l_c< l\le M,$$
in which $M$ is the truncation limit of $n$. The $\epsilon$ and $l_c$
have been taken to be $\epsilon={2 \over M^3}$ and $l_c=2 M^{3/4}$.
The pattern of $\sigma_l^{SV}$ as function of $l$ is shown in
Figure 1. 

In a viscous model for the astrophysical fluid simulations, the use of
artificial horizontal diffusion will not be needed, as it would always 
include the physical diffusion (for example, see \citet{fzlf1999}). 
The comparison between an inviscid model with artificial diffusion and
a viscous model without artificial diffusion can be done by using the
similar amount of artificial and physical diffusion in respective models. 
In order to derive an equivalent physical diffusion from the artificial
diffusion used in an inviscid model, a plausible experiment can be set up
as follows. By switching off all energy sources, such as kinetic energy 
from shear flow and the potential energy from effective gravity,
just the time-rate of the spread of a 2D Gaussian function in 
latitude-longitude direction can be studied in the presence of an
artificial diffusion. The horizontal diffusion, estimated from the
rate of spread of 2D Gaussian function, can be used as the amount of
physical diffusion in a viscous model without any artificial diffusion. 
The nonlinear evolution of a shallow-water instability, specifically
the spectral blocking issue and the quality of energy conservation from 
such a viscous model can be compared with that from an inviscid model
with artificial diffusion. A forthcoming paper will address this 
comparison quantitatively. 

However, one of the limitations of a single-layer shallow-water model,
in both viscous and inviscid cases, is that the horizontal velocities are
independent of depth -- only the vertical velocity varies with depth. 
Thus there are no viscous terms that involve radial second derivatives 
of horizontal velocities in the horizontal components of the momentum 
equations. In reality, a full 3D tachocline model will be a viscous model. 
So a multilayer shallow-water model will be needed to connect the layers 
through some diffusion in the radial direction, as has been done in the 
case of a full 3D, multilayer tachocline model by \citet{clz2008}. 
Alternatively, the effect of radial diffusion can be explored by using 
a Newton's cooling type formalism in which the radial diffusion terms 
can be converted into drag terms (for example, see \citet{dcg2004}). 
In future, development of such a model will also help investigate the 
effect of forcing the differential rotation at the top of the layer.  

\subsection{Adjusting the time step}

In order to increase the efficiency of the numerical code, we use
an adjustable time step which can be changed (up or down) during the 
the simulation. If the time step, $\Delta t$, is taken large, it may 
result in poor energy conservation, whereas if it is taken small, it will 
lead to longer runs that will take many steps and more time to compute. In 
the course of a computation, if the system is going through an interval  
near the peak of its potential or kinetic energy, then a somewhat lower 
time step will ensure better accuracy. At other times of the evolution,  
a larger time step can be taken, because that will speed up the 
computation. Therefore we have implemented a variable time step that is 
adjustable at each iteration. The scheme for adjusting time step
is described below.

We define $\epsilon$ as the fractional update of a variable-- the ratio
of the magnitude of the update of that variable and the magnitude of the variable
at that time step. When evaluating $\epsilon$ we exclude those
variables whose absolute magnitude is less than $10^{-4}$, because the
effect of those variables on the physical dynamics at that point in time
is very small compared to the normalized variables. Then we evaluate 
the maximum value of $\epsilon$ over all the updates at that 
time step. If the maximum in $\epsilon$ turns out to be less than 
$0.005$ then we scale up the next time step by a factor of  $1.02$ of 
current time step. On the other hand, if the maximum in $\epsilon$ is 
greater than $0.005$ then we scale down the time step by $0.8$ and 
reevaluate the evolution of the variables at that time step. In this way 
we limit the amount of maximum update of a variable, and hence the drastic
change in the system at a given time step, to achieve good energy 
conservation over long periods of time.

Values of scale-down and scale-up in the time-step are chosen
to be 20\% and 2\% respectively. These are somwhat arbitrary, but we have found
for our present calculations the amount of scale-down in the time-step
needs to be larger than that of scale-up. This is because
accuracy in the conservation of the energy deteriorates when the
system is evolving rapidly.   

\section{Results}

In order to present results and physical interpretations in a 
well-organized fashion, we divide this section into four  
subsections, which will successively present the convergence 
test related to the selection of longitudinal wavenumbers, a test for
total energy conservation and exchanges among different
energy reservoirs, nonlinear evolution of longitude-averaged flow and 
height profiles, and nonlinear evolution of flow accompanied by  
deformations of the fluid shell in latitude and longitude. 

\subsection{Convergence test for truncation in $m$}

For the technique described in detail in the appendix, the computation 
speed largely depends on the number of $l$'s and $m$'s we select in the basis 
functions. The total energy is defined in this system as 
$$ T_{en}=\int \int d\phi d\theta\cdot\left[(1+h) \left({u^2+v^2 \over 2}\right)
  +{G\cdot h\left(h+2\right) \over 2}\right], \quad\eqno(12)$$
in which $u$, $v$ and $h$ represent longitudinal velocity, latitudinal
velocity and height deformation.
Since the modes with longitudinal wavenumbers $m=1$ and 2 are the only
modes that are linearly unstable in the solar tachocline, we construct
our initial configuration of the flow and height-deformation by using
the modes with $m$ equal to $2$ at the most. 

We calculate the energy distribution at each longitudinal wavenumber $m$,
using the expression (12), for two $G$ values, namely $G=10$ and
$G=0.5$, and plot them in Figures 2(a) and (b) as functions of $m$. 
In both cases we have taken a pole-to-equator differential rotation 
amplitude of 24\% with $s_2=0.15$, $s_4=0.09$ in the expression (13). 
The total energy of the system for different wavenumbers $m=1$
through 10, plotted at the time $t=0$, $t=50$ and $t=90$ (respectively
by black, red and blue diamonds) show that the energy contribution falls 
off rapidly with $m$. At $t=0$ the total energy of the
system is contained in $m=0$ and $m=1$, because that is how we 
constructed the reference state. which suggests
that $m$ may be truncated at $6$ or $7$ with very little
loss of accuracy; this speeds up our computation enormously.

\clearpage
\begin{figure}[hbt]
\epsscale{0.75}
\plotone{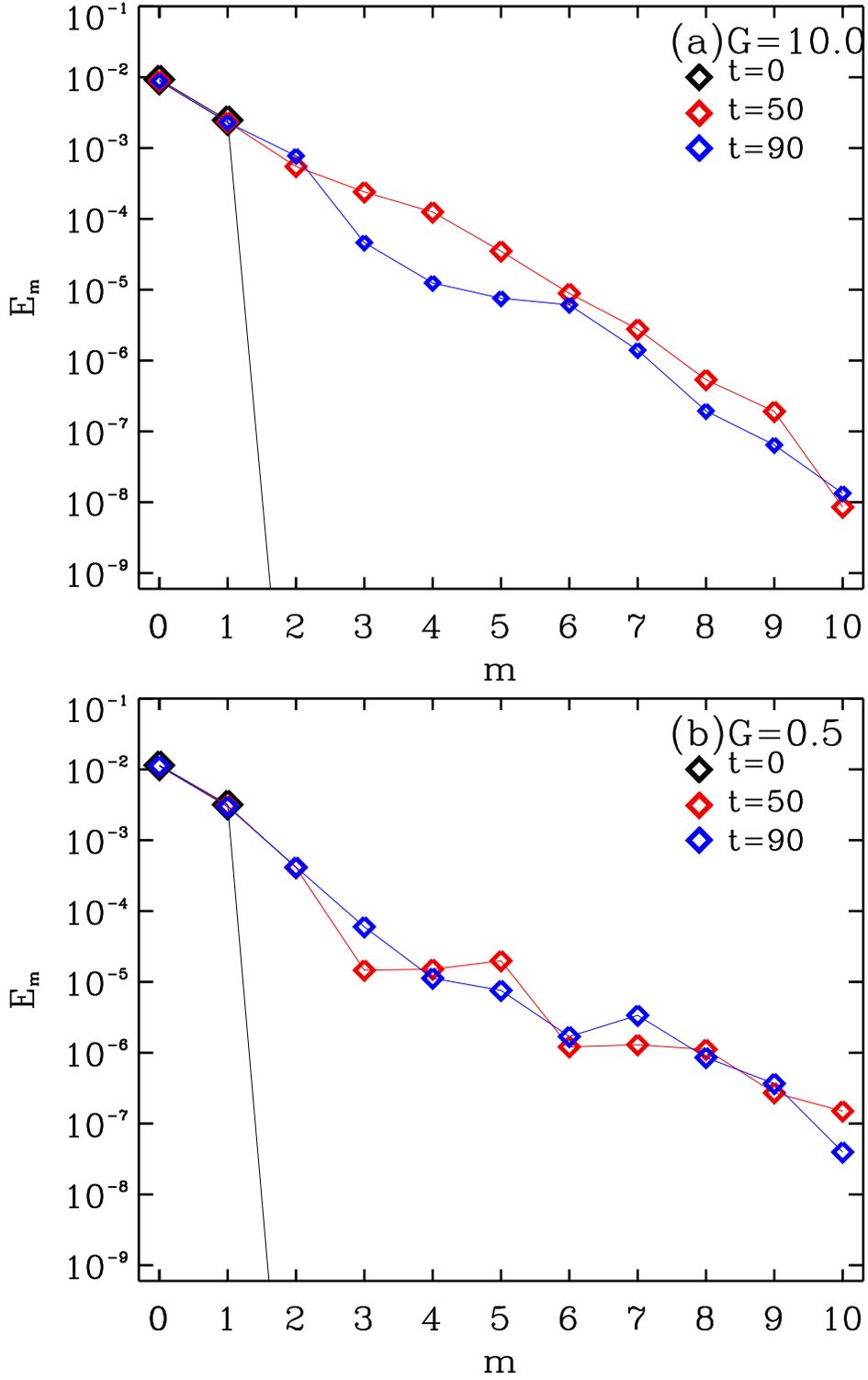}
\caption{Spectrum of total energy as function of longitudinal wavenumber $m$
at $t=0$ (black diamonds), 50 (red diamonds) and 90 (blue diamonds), for (a)
$G=10$ and (b) $G=0.5$. Note that dimensional time for $t=40$ and 90 are
respectively 4.8 and 10.8 months.}
\label{convergence}
\end{figure} 
\clearpage

\subsection{Evolution of energetics}

Because the fluid-shell system with latitudinal differential rotation
has its own inherent instability for low-order longitudinal wavenumber
modes, in the course of the nonlinear evolution of such a system its 
unstable modes grow by taking energy from the unperturbed system and 
thus modifying the system. At a certain point the reference-state 
system gets modified to the extent that 
it takes back energy from the perturbations. The cascading of energy to
higher longitudinal modes is not significant in this system
for $m$ below 6 or 7; instead a stable oscillation
of energy between the reference-state and perturbations develops. 
Figures 3(a) shows such an oscillation between the reference-state 
($m=0$) kinetic energy ($\bar{K}$) and the perturbation ($m>0$) kinetic 
energy ($K^{\prime}$), for a perturbation of about 40\% in the differential
rotation (angular measure, i.e. $u=\omega\cos\theta$), for the case
of $G=0.5$. Similar oscillations occur in $G=10$ case.

The same period and phase of oscillation occurs between the
reference state ($\bar P$) and perturbation ($P'$) potential
energies in the system, but with substantially smaller
amplitude than for the kinetic energy. We see that during this 
oscillation the total energy of the system remains virtually constant,
which shows that the numerical viscosity in the system is
extremely small. In fact, the total kinetic and total
potential energies are separately constant, so that the
energy exchanges are strictly between reference state
and perturbation energies of the same type.

\clearpage
\begin{figure}[hbt]
\epsscale{0.7}
\plotone{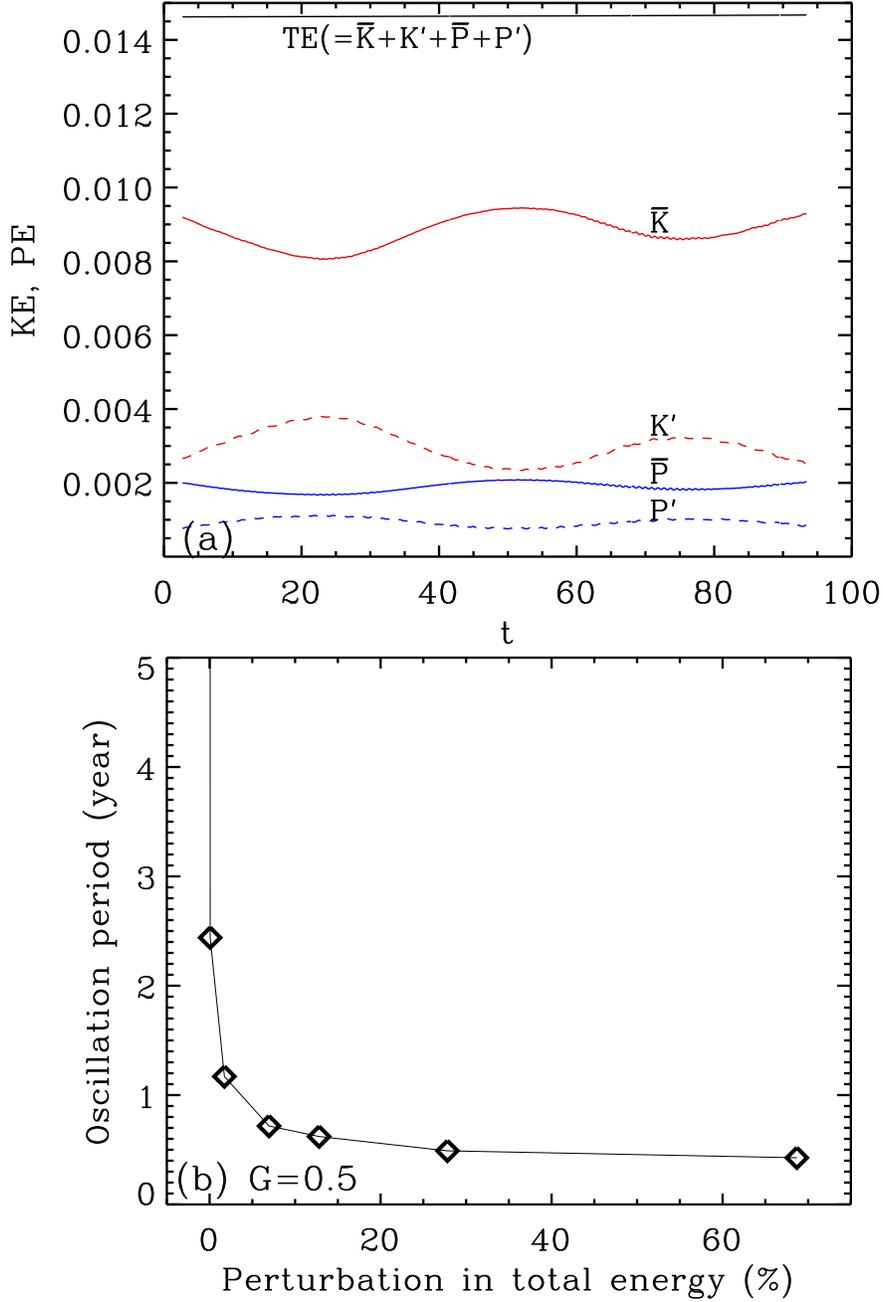}
\caption{Frame (a) shows the evolution of reference state kinetic energy, 
$\bar{K}$ (i.e., ${KE}_{m=0}$, plotted in red solid curve), potential energy, 
$\bar{P}$ (i.e., ${PE}_{m=0}$, blue solid curve), perturbation kinetic energy, 
$K'$ (i.e., ${KE}_{m>0}$, red dashed curve), and potential energy, $P'$ (i.e., 
${PE}_{m>0}$, blue dashed curve). The energy exchanges between $\bar{K}$ and 
$K'$, and between $\bar{P}$ and $P'$ shows an oscillatory pattern, the 
frequency of which depends on the amplitude of the perturbation, while total 
energy, plotted in solid black curve, is conserved. Frame (b) shows the 
oscillation period of energy exchange between the reference state and 
perturbation as function of perturbation energy measured as a fraction
of the reference state energy of the system.}
\label{energy-oscillation}
\end{figure} 
\clearpage

From equation (56) of Dikpati \& Gilman (2001) we can infer that 
this oscillation is driven by the time dependent Reynolds stress 
of the perturbations. Initially this stress extracts kinetic energy 
from the differential rotation, which is modified in such a way as 
to reduce the energy available for continued growth in the disturbance. 
But the system then 'overshoots' an equilibrium state, resulting at 
later phases in the oscillation in the perturbations giving back 
kinetic energy to the differential rotation. Then the perturbations 
are able to grow again, and the oscillation repeats. In \S3.4 we will 
show how the disturbance structure changes during the oscillation.

Figure 3(b) shows how the oscillation period depends on the initial
perturbation amplitude. The relationship is obviously not linear. The 
process that causes this behavior is also nonlinear. Physically what is 
happening is that as the initial perturbation amplitude is increased, 
energy is extracted from the differential rotation at a faster rate. 
This modifies the differential rotation profile faster to a form that 
is less unstable to the perturbation and subsequently to a form that
extracts energy from the perturbation. This makes the period shorter.
There is probably a limit to how short the period can be, as evidenced 
by the curve in Figure 3b approaching an asymptote in period that is
greater than zero.

Conversely for very low initial amplitude the period gets longer, but 
probably not to infinity because the initial growth is exponential 
(and the rate is independent of the small amplitude). It simply takes 
longer for the growing disturbance to reach a point in time at which 
the feedback from the modified differential rotation is strong enough 
to limit further growth.

\subsection{Nonlinear evolution of differential rotation -- formation
of high-latitude jets}

We have seen above that the kinetic and potential energies of the
nonlinear shallow water system go through a well defined oscillation.
How is this manifested in the changes in the differential rotation
profile? Figures 4a,b show its behavior for the same time period,
for both high $G$ (frame a) and low $G$ (frame b), for a 40\% initial
disturbance amplitude. Here is plotted the angular velocity in the shell
for five later times, compared with the initial profile (black curve).
All angular velocities are taken relative to the core rotation rate.

\clearpage
\begin{figure}[hbt]
\epsscale{0.85}
\plotone{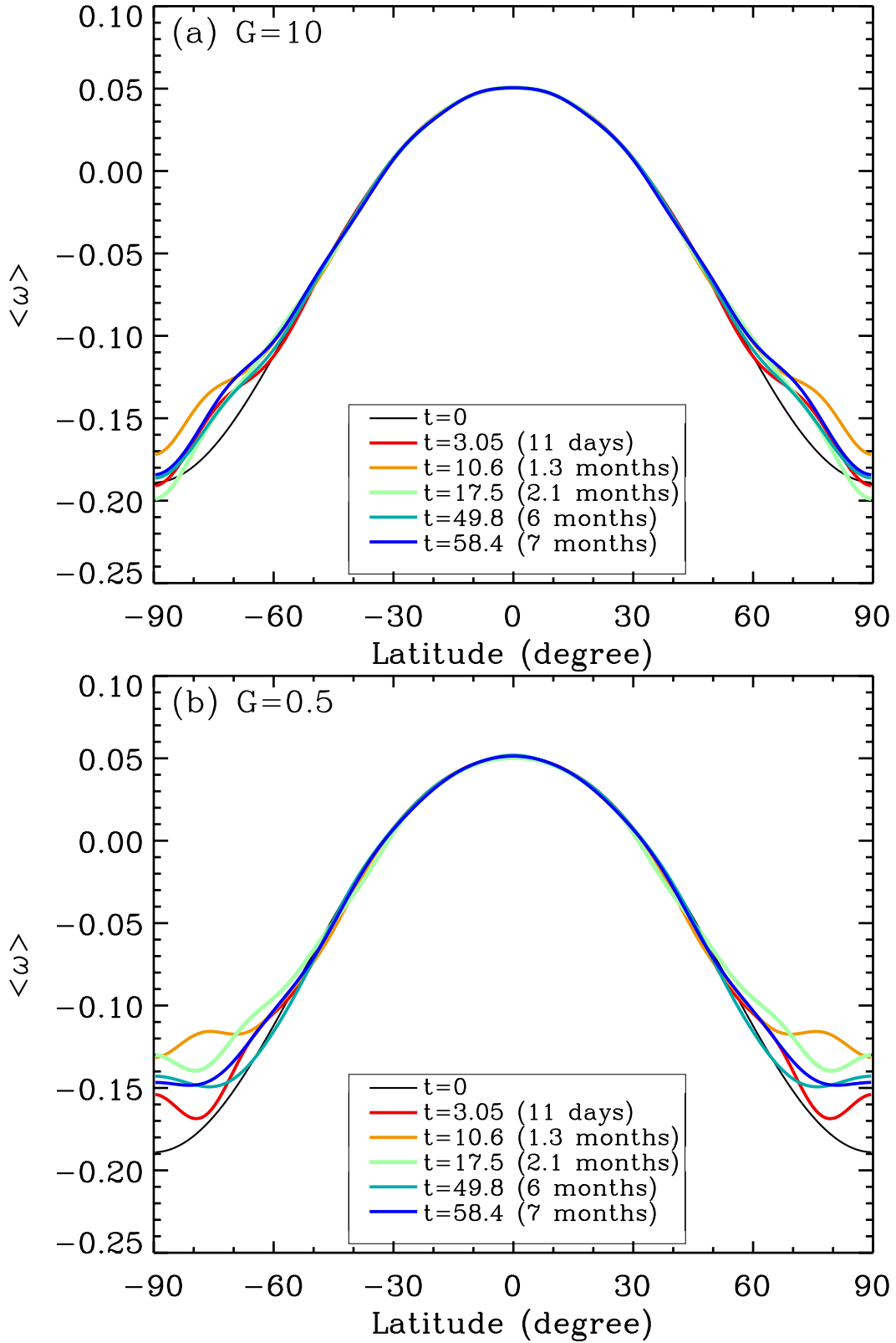}
\caption{Frame (a) shows, for a high effective gravity ($G=10$) snapshots 
of longitude-averaged differential rotation at several selected time 
during the course of its evolution. Frame (b) shows the same, but for
a low effective gravity ($G=0.5$).
}
\label{diff-rot}
\end{figure} 
\clearpage

We see in Figure 4 that the changes in angular velocity with time
are much larger in high than in low latitudes. This is because it
is angular momentum that is being transported in latitude by the
Reynolds stresses of the perturbations, and the moment of inertia
and volume of high latitudes is much smaller than that of low latitudes.
So depositing a given amount of angular momentum in high latitudes
causes the angular velocity to rise much more there than it falls
in low altitudes. It follows that if this instability is
active in the solar tachocline, the observed high latitude angular
velocity changes would be much larger than those in low latitudes.
To date, helioseismic measurements have not been able to
tell us whether this is so, because high latitudes are very hard
to observe.

For both high and low $G$, the peak high latitude angular velocity
is reached within a little over one month after the initial
disturbance is introduced. Thereafter it fluctuates, but about a
distinctly higher polar angular velocity than in the initial state.
The polar angular velocity is significantly higher for low $G$ than
high. We reason below that this is because with low $G$ it is easier
for the mass distribution in the shell to be rearranged by meridional
flow. The detailed time history of the amplitude of the polar
angular velocity does not precisely match the rather smooth
oscillation seen in Figure 3. This is not surprising, because
the time variation seen in Figure 3 is of global integrals of
the energy, while the polar rotation changes are for a very small
part of the volume of the global shell.

We can illustrate and explain the physical processes involved in
the evolution of the differential rotation by using equations (60)-(62)
of Dikpati \& Gilman (2001), which we repeat here, rearranged so that
only the time derivatives are on the left hand sides.
$${\partial \overline{u_2} \over \partial t} = 
F_u -\overline{v_2}
{\partial \over \partial\mu}\left(\omega_0 \thinspace (1-\mu^2)\right),
\quad\eqno(13)$$
$${\partial \overline{v_2} \over \partial t} = 
F_v -2 \thinspace \omega_0
\thinspace \overline{u_2}\thinspace \mu+ G \sqrt{1-\mu^2}
{\partial \overline{h_2} \over \partial\mu}, \quad\eqno(14)$$
$${\partial \overline{h_2} \over \partial t} =
F_h - {\partial \over \partial\mu}\left((1+\overline{h_0})\overline{v_2}
\sqrt{1-\mu^2})\right), \quad\eqno(15)$$

\noindent The forcing functions $F_u$, $F_v$ and $F_h$ are defined as,
$$F_u =-\overline{v'{\partial \over \partial\mu}
(u' \sqrt{1-\mu^2})},$$
$$F_v =-\overline{{u' \over \sqrt{1-\mu^2}}\left({\partial v' \over
\partial\lambda} - \sqrt{1-\mu^2} {\partial \over \partial\mu}
(u' \sqrt{1-\mu^2})\right)} -\sqrt{1-\mu^2}{\partial \over \partial\mu}
\left({\overline{u'^2}+\overline{v'^2} \over 2 } \right),$$
$$F_h =-{\partial \over \partial\mu}
(\overline{h'v'} \sqrt{1-\mu^2}).$$

Here the variables are $\overline{ u_2}$, $\overline{v_2}$ and 
$\overline{h_2}$, which represent respectively the departures of linear 
rotational velocity, meridional flow, and layer thickness from their 
initial values. $\omega_0$ is the angular velocity of the core, and 
$\mu$ is cosine of the latitude. From above definitions, one can see
that $F_u$, $F_v$ and $F_h$ are quadratic functions that contain the 
covariances or stresses from the perturbation variables $u, v, h$. Equations 
(13) and (14) are, respectively, the equations of motion for differential
rotation and meridional flow, and Equation (15) is the mass continuity
equation, all for the longitudinal wavenumber $m=0$ variables.

Figure 5 displays profiles of layer thickness, meridional flow
and angular velocity at the same times as shown in Figure 4b.
Thickness (solid purple curve) and latitudinal flow (solid green curve)
are shown in frames (a) through (e), along with the initial thickness
(dashed purple curve) for reference. Note that initial meridional flow 
(dashed green curve) is zero.
The differential rotation angular velocity (thick black curve) is
shown in frames (f)-(j), together with the initial differential rotation
(thin curve).

If we compare the thickness and latitudinal flow profiles as time advances, 
we can see clearly that initially the thickness shrinks near the poles 
but later rises again as the oscillation progresses.
Consistently, the meridional flow is away from the poles while the
thickness is declining there, and toward the poles while the polar
thickness is rebounding. The same relationship between meridional
flow and thickness, dictated by mass conservation, can be seen at all 
other latitudes. When the polar thickness is declining, the polar
angular velocity is increasing.

The physical processes that govern these correlated changes among
thickness, meridional flow and differential rotation are contained
in Equations (13)-(15) above (see \citet{dg01}). In brief,
what happens during the nonlinear oscillation is that, through $F_u$
(Equation 13) the perturbation Reynolds stress transports angular 
momentum toward the poles to cause a 'spin-up', or higher linear 
and angular velocity there. In the meridional flow Equation (14), this
'spin up' leads to a negative (equatorward) Coriolis force that pushes
fluid away from the pole toward the equator. This mass flow then
produces a positive latitudinal pressure gradient force, which is 
expressed as a thickness gradient in a shallow-water model. This
force starts to counterbalance the Coriolis force.

\clearpage
\begin{figure}[hbt]
\epsscale{0.65}
\plotone{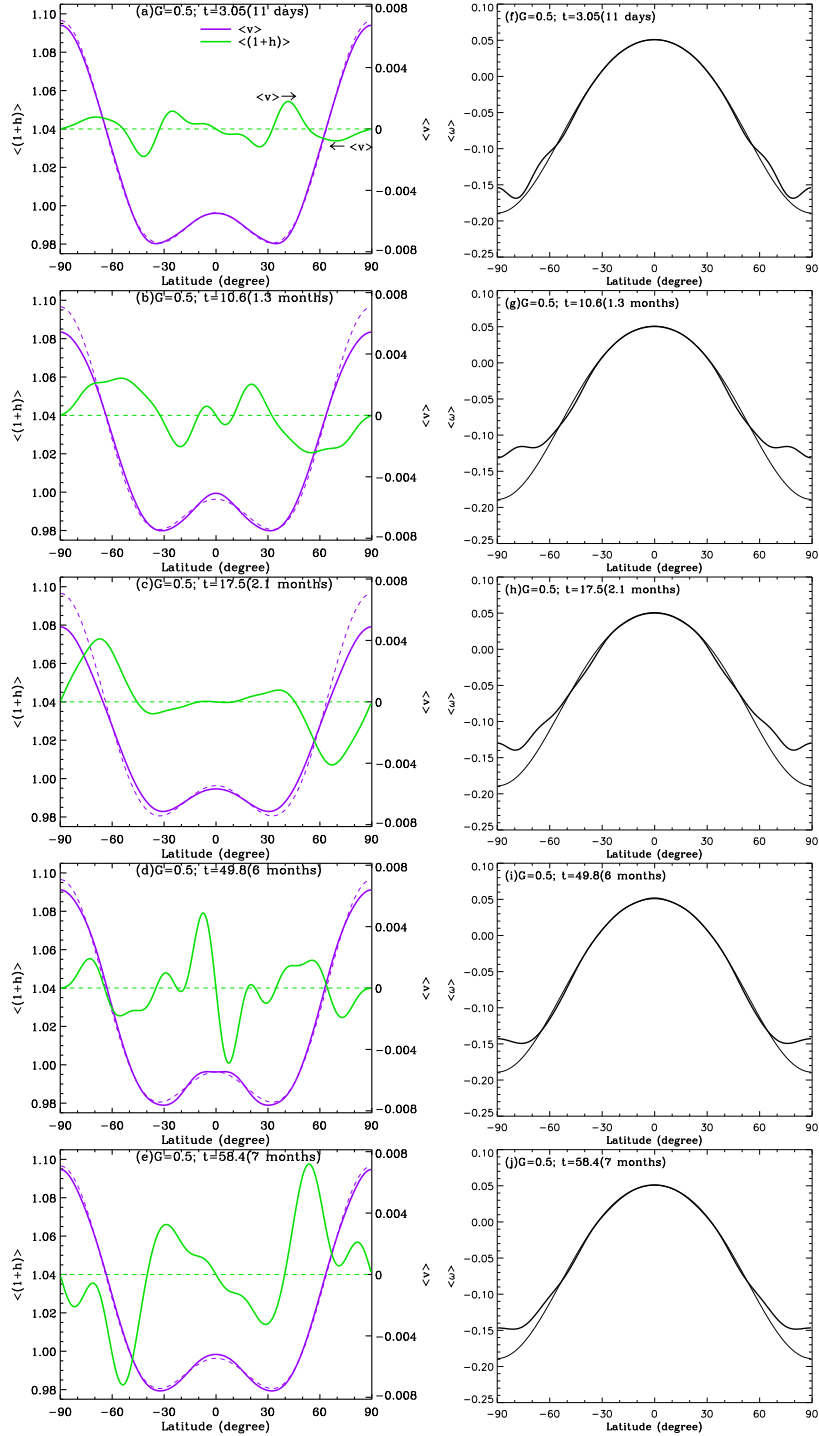}
\caption{Frames (a-e) show, for $G=0.5$, five snapshots of longitude-averaged
height deformation (solid purple curves) for the same selected times as 
in Figure 4(b), and corresponding latitudinal flow (solid green curves) 
have been superimposed on that. Dashed purple and green curves in each
frame respectively present height deformation and latitudinal flow
at $t=0$. Solid black curves in frames (f-j) show longitude-averaged 
differential rotation, for the same selected times, with a superimposed
dashed curve for $t=0$. The right column is created by separating 
five curves of Figure 4b.
}
\label{dynamics}
\end{figure} 
\clearpage

This rebalancing of forces will reduce the equatorward meridional 
flow, but in our nearly dissipation free system, it overshoots, causing
the meridional flow to reverse and rebuild the polar thickness.
At the same time, the changed differential rotation profile is
less unstable to the perturbations, or even stable to them,
so the Reynolds stresses get much weaker, and the polar latitudes 
spin down again. Throughout this sequence of changes, the mass 
continuity Equation (15) ensures that the changes in thickness are 
consistent with the changes in meridional flow.

\subsection{Evolution of disturbance planforms}

We can better understand the nonlinear dynamics of
the shallow water system by examining the planforms
(longitude-latitude) of the flow and thickness as they
evolve. Figure 6 displays these quantities for the same
times as in Figures 4b and 5 (low $G$, 0.5). The left hand
column displays the total flow and total thickness fields
(blue is thinest, red thickest); the right hand column
shows all of the $m>0$ fields, after subtracting out
the axisymmetric parts of the differential rotation, 
meridional flow and axisymmetric thickness. 

From both columns we can see that the $m=1$ mode dominates in
the flow and thickness patterns, leading to a 'meandering' of
the E-W flow. As time progresses, more structure and somewhat 
finer spatial scales develop, as we should expect from the behavior 
of the energy spectrum shown in Figure 2. We can also see from both
columns that the flow is predominantly geostrophic, that is
in the North there is counterclockwise circulation around
the blue (thinest) areas, clockwise around the red (thickest)
areas; the reverse is true in the South. Since the pressure
in the shallow water system is hydrostatic, it is proportional
to the thickness, so the counterclockwise flow in the North is around
low pressure, the clockwise flow around high pressure.

From the right hand column we can see that in the first 1-2 months the
disturbances are transporting angular momentum toward the poles to spin 
them up, as seen in Figure 4b. The velocity and thickness patterns
are tilted with respect to the E-W direction in such a way that
the fluid elements moving toward the east in the disturbance are
also moving toward the poles, while those moving toward the west
(relative to the rotating reference frame) are moving toward the
equator. This means there is a correlation between the two velocity
components that is the Reynolds stress, which is carrying out
the poleward angular momentum transport.

\clearpage
\begin{figure}[hbt]
\epsscale{0.8}
\plotone{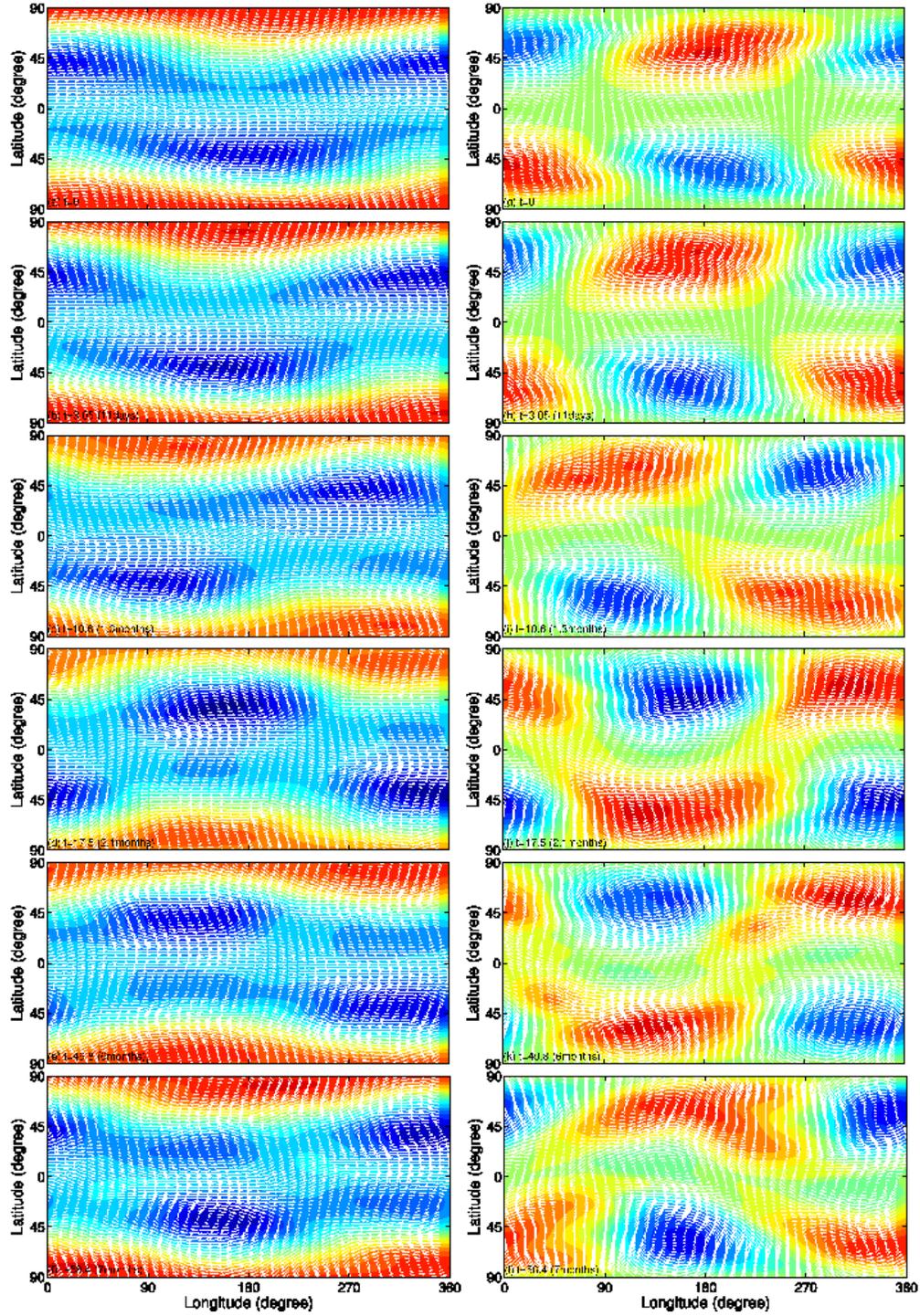}
\caption{Frames (a-f) show, for $G=0.5$, planforms of flow (arrow vector)
and height (color map) for five selected times as in Figure 5. Frames
(g-l) show the perturbation part of flow and height. Red represents
swelling and blue depression.
}
\label{planforms_unstable}
\end{figure} 
\clearpage

\begin{figure}[hbt]
\epsscale{0.8}
\plotone{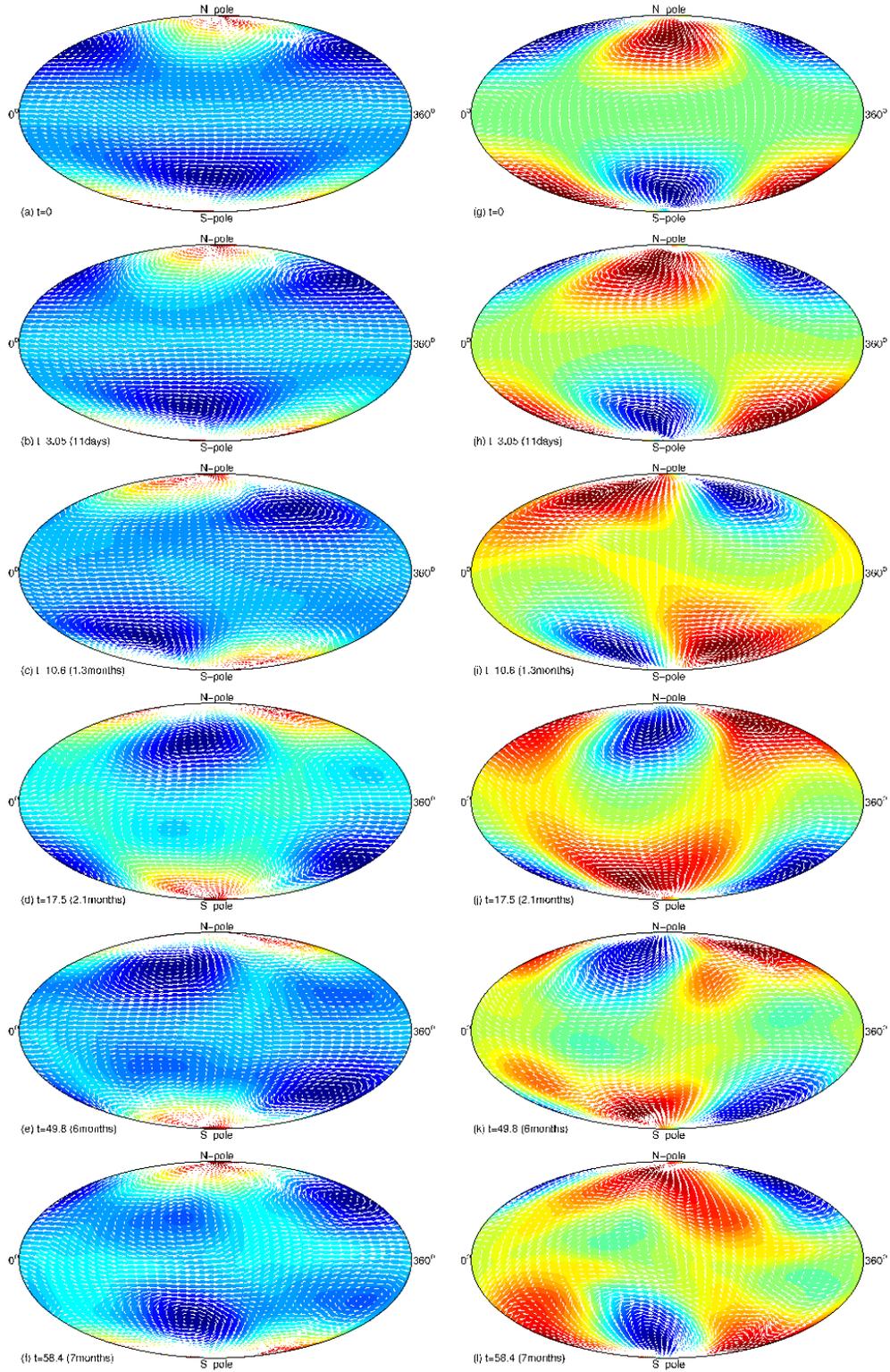}
\caption{Same as in Figure 6, but in Mollweide projection.}
\label{planforms_unstable_mollweide}
\end{figure} 
\clearpage

By 2.1 months, these tilts are virtually gone, indicating angular
momentum is no longer being transported toward the poles. This is 
consistent with the profiles of differential rotation shown in Figure 4,
in which the maximum amount of polar spin-up is reached between 1.3 and 
2.1 months in these selected cases. By six months, the perturbation patterns
have become more complex, and it is difficult to discern which way
the angular momentum transport is going. But from Figure 3 there is 
still a pronounced amplitude oscillation between the perturbations
and the differential rotation and meridional flow. In this nearly
dissipation-free system there does not appear to be an absolutely
steady nonlinear state. Instead there is a dynamical equilibrium 
with continuous evolution of all flow patterns 
within a nonlinear regime in which the high latitude rotation rate, 
while fluctuating, is almost always faster than in the initial state.

In Figure 7, we repeat the twelve frames (Figure 6a-l) in Mollweide 
projection. This Figure captures a more realistic look of the global
disturbance patterns, particularly at and near the poles. In latitude-longitude
plots, the single point at each pole is spread out over $360^{\circ}$
longitude. Figure 7 reveals all the features of the disturbances seen
in Figure 6, but also reveals more clearly that the flow spreads over 
both hemispheres crossing the equator. 

\subsection{Is a linearly stable profile nonlinearly unstable?}

To illustrate further the range of nonlinear behavior of the
shallow water system, we have performed an additional numerical 
experiment in which we have chosen a combination of differential
rotation ($s=0.15$) and effective gravity ($G=10$) that we know 
from linear theory is stable to perturbations, and initially 
perturbed the system with a disturbance calculated for an 
unstable case. The amplitude of perturbation used is 20\% of the 
reference state differential rotation. Shown in
Figure 8 is the resulting differential rotation evolution
sampled at the same times as in previous cases. Here we see that
in the first few days of the integration, the disturbance
starts to spin up the poles (red curve) because the Reynolds stress 
in it transports angular momentum toward the poles. But this trend is
very short-lived, as the stable differential rotation modifies the 
disturbance, changing the sign of its Reynolds stress by changing the
sense of tilt of the perturbation velocities with respect to the 
E-W direction. Consequently for the next two months of the integration,
the poles actually spin down to lower rotation than they had initially,
(light green curve) while the kinetic energy of the differential rotation 
is increased at the expense of the energy in the disturbance. But this, 
too, is a transient, and by seven months the rotation at
high latitudes has returned to very close to its initial profile
(dark blue curve). 

\begin{figure}[hbt]
\epsscale{1.0}
\plotone{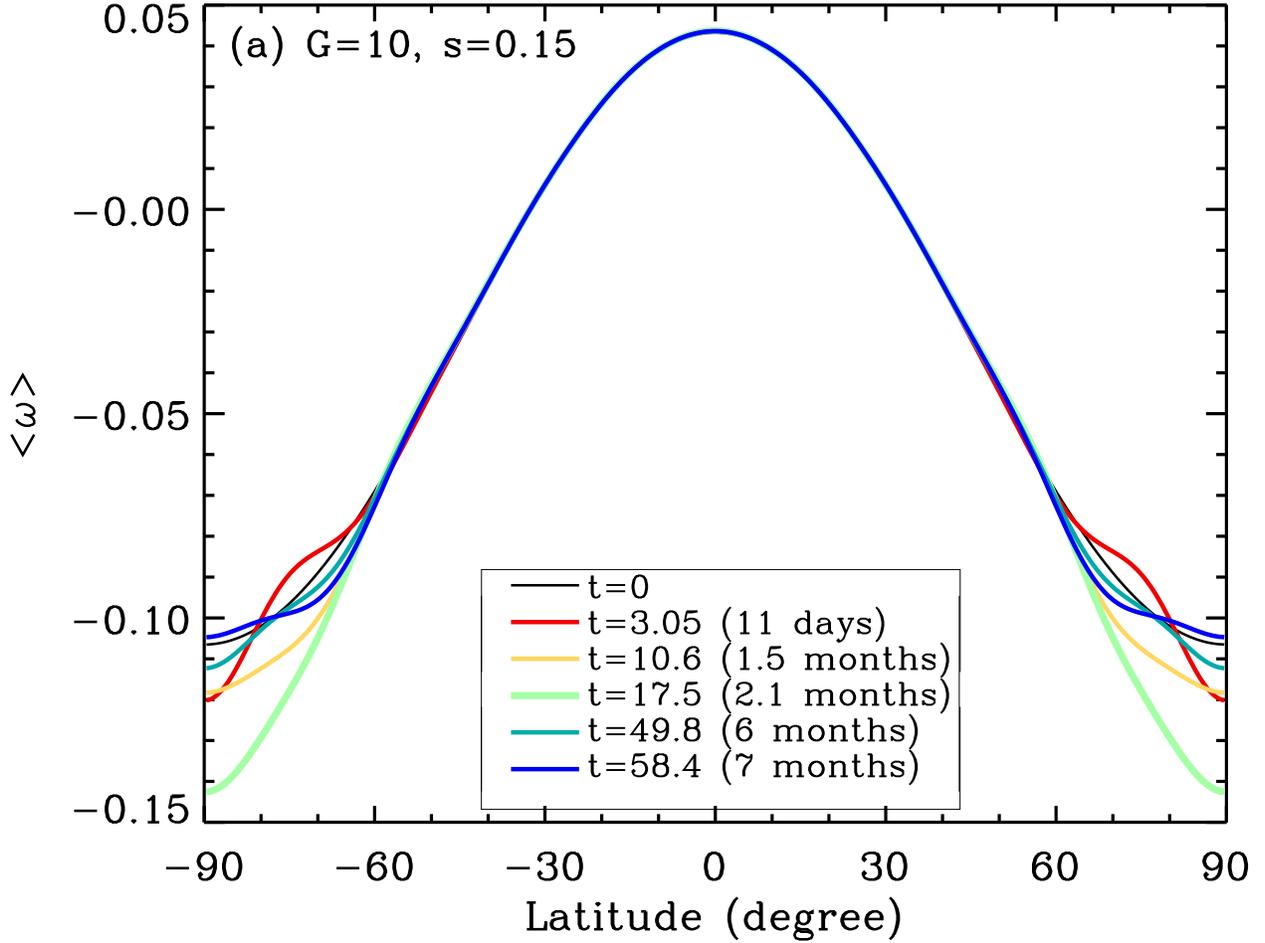}
\caption{Response of a linearly stable solar differential rotation 
(with $s=0.15$ and for $G=10$) to perturbation. Five snapshots present
the evolution of longitude-averaged differential rotation for five 
selected times as in Figure 5. 
}
\label{stable-case}
\end{figure} 

The details of how the disturbance evolves with time is shown in 
Figure 9, which gives longitude-latitude planforms of velocity
and shell thickness. The three frames are for, respectively,
$t=0, 10.6$ and 49.8 (0, 1.3 months, and 6 months). We see that in 
this succession of frames the tilt of the velocity vectors with 
respect to a latitude circle changes orientation twice, as the 
associated Reynolds stress changes sign twice, signifying first 
poleward, then equatorward, then poleward momentum transport again. 
This is what causes the poles to spin up, then spin down, then up again. 

\clearpage
\begin{figure}[hbt]
\epsscale{0.8}
\plotone{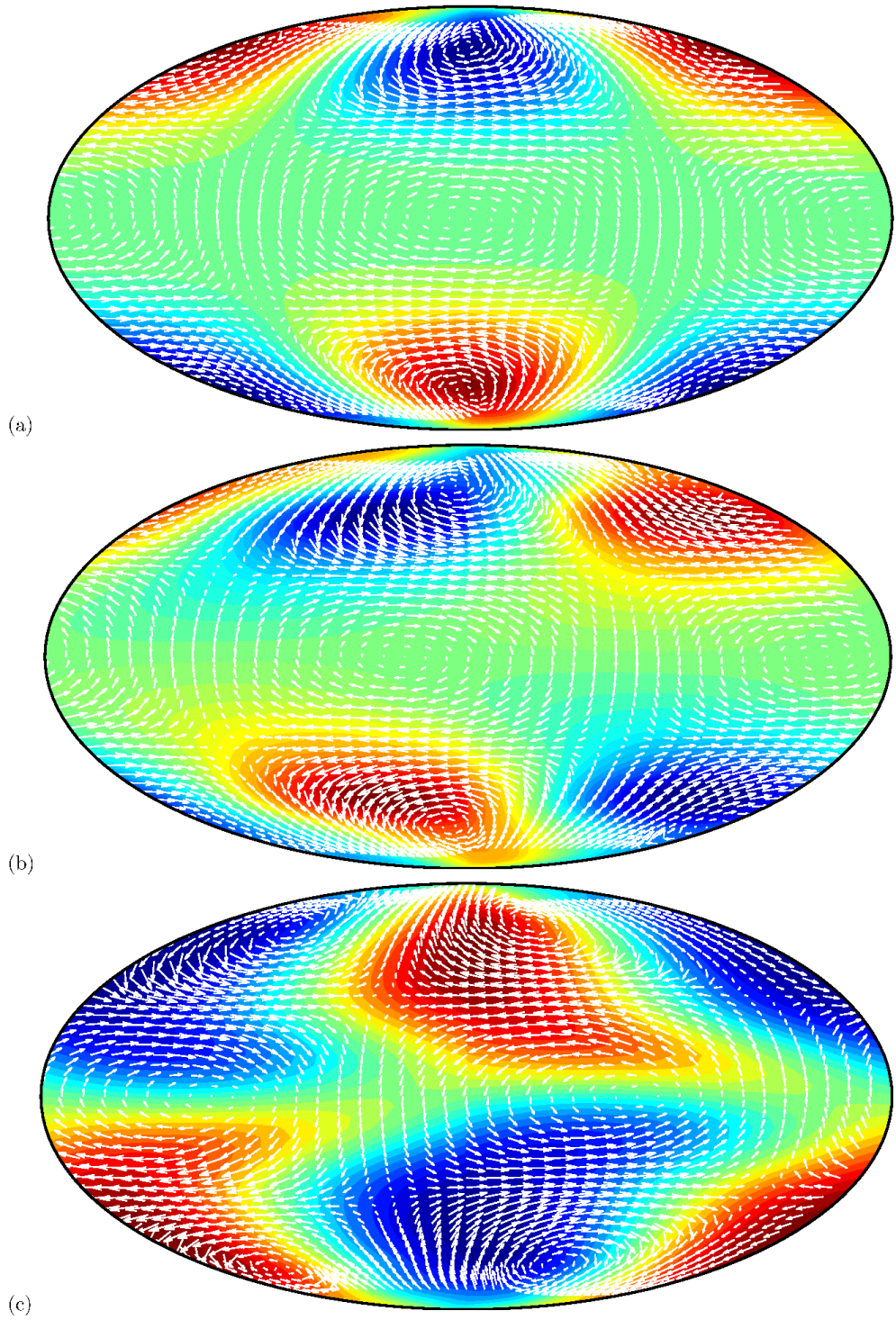}
\caption{
Frames (a-c) show, for $s=0.15$ and $G=10$, planforms of flow (arrow vector)
and height (color map) in Mollweide projection, for three selected times, 
$t=0$, 1.3 months and 6 months. 
}
\label{planform_stable}
\end{figure} 
\clearpage

Since linear studies \citep{dg01} indicate that the low differential 
amplitudes ($s\lesssim 0.17$) are stable in the radiative tachocline with 
high $G$, we considered a case with $s=0.15$ and $G=10$ in this subsection.
In linear studies shallow-water gravity modes were extracted out from the 
calculation and only shear modes were explored. However, we have no such 
freedom in the case of nonlinear evolution of a shear profile in the
radiative tachocline fluid shell with high $G$. Consequently we see
the appearance of gravity modes in the evolution discussed in Figures
8 and 9. The generation and evolution of gravity modes is a vast 
subject of research itself. A forthcoming paper will address the 
generation of shallow-water gravity modes due to nonadjustment of 
geostrophic balance when the effective gravity of the system is high.

\section{Concluding Comments}

We have presented above the first fully nonlinear 'shallow water' numerical 
model to be applied to the Sun. It is intended
for use in understanding the global dynamics of solar and stellar tachoclines.
It provides a beginning and template for later models that will include global
MHD effects; solar tachoclines most likely contain strong
global scale magnetic fields, particularly toroidal fields.

This spectral model needs only extremely low numerical diffusion to keep 
the solutions well behaved, so it is particularly well suited to simulation 
of strongly nonlinear time dependent hydrodynamics and MHD of the Sun. The 
equations conserve energy with extremely high accuracy ($\sim 0.001$\%), so 
that very long simulations are possible without significant spurious build-up 
of kinetic or potential energy at small scales. Substantial increases in 
both computational efficiency and accuracy are obtained by use of a mechanism 
for adjusting the time step during a simulation.

Since the model does include the radial dimension in a simple way, by
means of a deforming top surface, it contains gravity wave modes as well as 
shearing instabilities of the differential rotation. It also allows for
a simplified form of meridional flow that is coupled to and causes the changes
in the thickness of the spherical shell. Thus it is possible with this model
to study rather complex fluid dynamics, including interplay between the
relatively low frequency global shearing flows and instabilities, and the
higher frequency gravity waves made possible by the deforming top boundary.

The low dissipation in the model is of particular value for simulating the
global dynamics of the 'radiative' tachocline that is found below the 
overshooting layer of the convection zone above. But it is equally valuable
for simulations applied to the 'overshoot' tachocline, whose differential
rotation is maintained by downward diffusion from the convection zone.

Our first results from this nonlinear model focus primarily on the nonlinear
interactions between unstable global perturbations previously found from
linear analyses and the differential rotation that gives rise to them. We
show that the nonlinear dynamics result in a 'spin-up' of polar regions
due to the Reynolds stress associated with the growing disturbances, and also
substantial nonlinear oscillations between disturbance and differential
rotation kinetic energy. The oscillation occurs because as the disturbances
extract energy and transport momentum to high latitudes, the differential
rotation profile changes in such a way as to reduce and even shut off the 
instability. At least temporarily there can occur a reversal in direction
of angular momentum transport, which rebuilds energy in the differential 
rotation. The simulations go through a sequence of these oscillations without
settling to a steady state. This is because the system is virtually
dissipation-free due to the implementation of numerical scheme with very
little spectral viscosity. Throughout this evolution,
the longitude-latitude planform of the disturbances on the differential
rotation, as well as the total flow, undergoes considerable evolution as well
as propagation in longitude.

The changes in angular velocity with time in polar latitudes are much larger 
than the compensating changes in low latitudes, because the moment of inertia 
and volume of polar latitudes is much smaller than in low and mid-latitudes. 
The period of the nonlinear oscillation is inversely proportional to the 
amplitude of the initial perturbations, apparently with finite asymptotic 
periods for both very large and very small initial disturbance energies.
Throughout these oscillations the disturbance kinetic energy remains
confined primarily to the lowest longitudinal wave numbers, $m=1,2$, or
the largest global scales.

The simulations contain gravity waves of significant amplitude particularly
when the effective gravity ($G$) of the system is high;
their role will be addressed in a later paper. The solutions also contain
kinetic helicity due to the correlation between the local radial component
of vorticity and the radial motion associated with changes in the thickness
of the shell. Kinetic helicity is of particular importance for the solar
dynamo, and will also be studied in more detail in a later paper. In the Sun
the differential rotation of the tachocline is imposed from the convection zone
above. We will generalize our model to include this forcing in later studies.
This will add still another time scale to the problem, namely the time it takes
for the top forcing to restore the differential rotation. 

All of the additional studies just described are for the hydrodynamic model 
we have already built. The major step beyond that system is to include global 
MHD processes, which will without doubt substantially modify the hydrodynamic 
processes we have studied or will study soon. A still later step will be to 
couple this model to a global 3D flux transport dynamo model for the Sun, 
that will be capable of addressing the question of the origin of active 
longitudes \citep{dg05} as well as how such longitudinally dependent features 
and processes contribute the origin and evolution of solar cycles.
The coupling would involve flow generated in the tachocline reaching into the
convection zone above and generating 3D MHD induction effects. Thus there is
a need to develop a viscous shallow-water model in order to be able to couple
the tachocline instabilities with a global, flux-transport dynamo model in 
the solar convective envelope. In addition, it is necessary for the magnetic 
fields in the tachocline to connect with that in the convection zone
above, continuously through the interface between the tachocline and the 
convection zone. To implement such a connection among magnetic fields 
at different layers, through the interfaces among the layers, we would
follow the way as described in the figure 1 and equation (22) of \citet{clz2008}.

\acknowledgements 

We thank Phil Judge for reviewing the entire manuscript and for his helpful 
comments. We extend our thanks to Yuhong Fan for helpful discussions on 
artificial viscosity in an inviscid model versus viscous model with physical
diffusion, to Peter Gilman for many helpful discussions on the topic of this 
paper, and to an anonymous reviewer for his/her constructive comments and 
suggestions, which have helped improve this paper. This work is partially 
supported by NASA's Living With a Star program through the grant NNX08AQ34G. 
The National Center for the Atmospheric Research is sponsored by the National 
Science Foundation.

\section{Appendix: Solution Technique}

In the case of primitive equations such as equations (8)-(10), the vector 
field (flow) and the scalar field (height) need to be expanded respectively 
in terms of vector and scalar spherical harmonics, in order to handle the 
pole problem. Pure scalar harmonics can also be used by converting all 
the vector fields into scalar fields, but at the expense of raising the 
order of the equations. 

Following the solution method of \citet{swarzt96} (see method 2 there),
the spectral decomposition of scalar $h$ can be expressed as
$$ h(\phi,\lambda) = \sum_{l=0}^N
{\sum_{m=0}^l}^\star P_{lm}\left(a_r^{lm}\cos{m\lambda}+a_i^{lm}
\sin{m\lambda}\right), \quad\eqno(A1)$$
where
$$
\sqrt{\alpha_{lm}}={2l+1 \over 2\pi} {\left(l-m\right)! \over \left(l+m\right)!},
\quad\eqno(A2)$$
and $P_{lm}$ is the Associated Legendre polynomial of order 
$(l,m)$. The $\star$ in the $m$ summation implies a factor of $\frac{1}{2}$ 
is multiplied when $m$ equals 0. The coefficients $a_r^{lm}$ and $a_i^{lm}$
can be obtained from the inverse transform as 
$$
a_r^{lm}=\int_{-{\pi \over 2}}^{\pi \over 2}
d\phi P_{lm}\left(\phi\right)\cos{\phi}
\int_{0}^{2\pi} d\lambda\cos{m\lambda}h\left(\phi,\lambda\right),
\quad\eqno(A3)$$
$$
a_i^{lm}=\int_{-{\pi \over 2}}^{\pi \over 2}
d\phi P_{lm}\left(\phi\right)\cos{\phi}
\int_{0}^{2\pi} d\lambda\sin{m\lambda}h\left(\phi,\lambda\right).
\quad\eqno(A4)$$

In order to express the vector fields ($u$ and $v$), we define the 
vector spherical harmonic components as

$$ \sqrt{l\left(l+1\right)}V_{lm}\left(\phi\right) = {dP_{lm} \over d\phi}
=\frac{1}{2}\left\{(l+m)(l-m+1)P_{l(m-1)}-P_{l(m+1)}\right\}, \quad\eqno(A5)$$
$$ \sqrt{l\left(l+1\right)}W_{lm}\left(\phi\right) = {m P_{lm} \over
\cos{\phi}} = {\left(-1\right) \over 2}\left\{P_{(l-1)(m+1)}+(l+m)(l+m-1)
P_{(l-1)(m-1)}\right\}. \quad\eqno(A6)$$

\noindent The unit vectors can be constructed as
$$
\begin{array}{ccc}
B_{lm}=\left(\begin{array}{c}iW_{lm}\\V_{lm}\end{array}\right)e^{im\lambda}
&,&C_{lm}=\left(\begin{array}{c}-V_{lm}\\iW_{lm}\end{array}\right)
e^{im\lambda}
\end{array}
\quad\eqno(A7)
$$

\noindent It is easy to check that $B_{lm}$ and $C_{lm}$ satisfy the 
following orthogonality relations

$$
\left(B_{ij},C_{lm}\right) = 0, \quad\eqno(A8a)$$
$$
\left(B_{ij},B_{lm}\right)=\left(C_{ij},C_{lm}\right)=\alpha_{lm}
\cdot\delta_{il}\cdot\delta_{jm}, \quad\eqno(A8b)$$

\noindent in which $\delta_{ij}$ is Kr\"onecker-$\delta$.
The velocity components can now be expressed in terms of vector spherical
harmonics as

$$
u\left(\phi,\lambda\right) = 
\sum_{l=0}^N{\sum_{m=0}^l}^\prime\left\{W_{lm}
\left(b_i^{lm}\cos{m\lambda} - b_r^{lm}\sin{m\lambda}\right) \right.$$
$$ +\left.V_{lm}\left(c_r^{lm}\cos{m\lambda} + c_i^{lm}\sin{m\lambda}
\right)\right\}, \quad\eqno(A9)$$

\clearpage

$$
v\left(\phi,\lambda\right) = 
\sum_{l=0}^N{\sum_{m=0}^l}^\prime\left\{V_{lm} \left(b_r^{lm}\cos{m\lambda} 
+ b_i^{lm}\sin{m\lambda}\right) \right. $$
$$ +\left.W_{lm}\left(-c_i^{lm}\cos{m\lambda} + c_r^{lm}\sin{m\lambda}
\right)\right\}. \quad\eqno(A10)$$

The coefficients, $b_r^{lm}, b_i^{lm}, c_r^{lm}, c_i^{lm}$, can be evaluated
using the following inverse transforms:

$$
b_r^{lm}={1 \over \sqrt{\alpha_{lm}}}
\left\{\int_{-{\pi \over 2}}^{{\pi \over 2}}d\phi\,\cos{\phi} V_{lm}
\int_{0}^{2\pi}d\lambda\, v\cos{m\lambda}-\int_{-{\pi \over 2}}^{{\pi \over 2}}
d\phi\,\cos{\phi}\,W_{lm}\int_{0}^{2\pi}d\lambda\, u\sin{m\lambda}\right\},
\quad\eqno(A11)$$

$$
b_i^{lm}={1 \over \sqrt{\alpha_{lm}}}
\left\{\int_{-{\pi \over 2}}^{{\pi \over 2}}d\phi\,\cos{\phi}\, W_{lm}
\int_{0}^{2\pi}d\lambda\, u\cos{m\lambda}+\int_{-{\pi \over 2}}^{{\pi \over 2}}
d\phi\,\cos{\phi}\,V_{lm}\int_{0}^{2\pi}d\lambda\, v\sin{m\lambda}\right\},
\quad\eqno(A12)$$

$$
c_r^{lm}={1 \over \sqrt{\alpha_{lm}}}
\left\{\int_{-{\pi \over 2}}^{{\pi \over 2}}d\phi\,\cos{\phi}\, V_{nm}
\int_{0}^{2\pi}d\lambda\, u\cos{m\lambda}+\int_{-{\pi \over 2}}^{{\pi \over 2}}
d\phi\,\cos{\phi}\,W_{lm}\int_{0}^{2\pi}d\lambda\, v\sin{m\lambda}\right\},
\quad\eqno(A13)$$

$$
c_i^{lm}=-{1 \over \sqrt{\alpha_{lm}}}
\left\{\int_{-{\pi \over 2}}^{{\pi \over 2}}d\phi\,\cos{\phi}\,W_{lm}
\int_{0}^{2\pi}d\lambda\, v\cos{m\lambda}
+\int_{-{\pi \over 2}}^{{\pi \over 2}} d\phi\,\cos{\phi}\,V_{lm}
\int_{0}^{2\pi}d\lambda\, u\sin{m\lambda}\right\}.
\quad\eqno(A14)$$

Substituting (A9-A14) in equations (7a) and (7b) and performing a few steps
of algebra (see equations A.14 and A.15 of \citet{swarzt96} for detailed 
derivation), we can obtain the following expressions for the variables 
$\zeta$ and $\delta$, which no longer contain the unbounded terms that 
cause the pole problem in spherical polar coordinate system:

$$ \zeta=\sum_{l=0}^N {\sum_{m=0}^l}^\star
\sqrt{l\left(l+1\right)} P_{lm}\left(c_r^{lm}\cos{m\lambda}+c_i^{lm}
\sin{m\lambda}\right),\quad\eqno(A15)$$

$$ \delta=\sum_{l=0}^N {\sum_{m=0}^l}^\star
\sqrt{l\left(l+1\right)} P_{lm}\left(b_r^{lm}\cos{m\lambda}+b_i^{lm}
\sin{m\lambda}\right). \quad\eqno(A16)$$

\noindent Similarly following \citet{swarzt96} (see his equations A.16
and A.17) the expression for the gradient of the scalar variable, 
$h$, can be given as

$${1 \over \cos\phi}{\partial h \over \partial\lambda}=
\sum_{l=0}^N {\sum_{m=0}^l}^\star
\sqrt{l\left(l+1\right)}W_{lm} \left(a_i^{lm}\cos{m\lambda}-a_r^{lm}
\sin{m\lambda}\right),\quad\eqno(A17)$$

$${\partial h \over \partial\phi}=
\sum_{l=0}^N {\sum_{m=0}^l}^\star
\sqrt{l\left(l+1\right)}V_{lm} \left(a_r^{lm}\cos{m\lambda}
+a_i^{lm}\sin{m\lambda}\right). \quad\eqno(A18)$$

By grouping the derivative and algebraic terms and using the compact
notations, defined as

$$\Xi_u = (\zeta + 2\omega_c\sin\phi)v, \quad\eqno(A19)$$
%and 
$$\Xi_v = -(\zeta + 2\omega_c\sin\phi)u, \quad\eqno(A20)$$ 
the time-evolution equations for $u$ and $v$ can be written as

$${\partial u \over \partial t} = \Xi_u -{1 \over \cos\phi}
{\partial \over \partial\lambda}\left[\left({u^2+v^2 \over 2}\right)
+Gh\right], \quad\eqno(A21)$$

$${\partial v \over \partial t} = \Xi_v -{\partial \over \partial\phi}
\left[\left({u^2+v^2 \over 2}\right)+Gh\right]. \quad\eqno(A22)$$

Note that $\Xi_u$ and $\Xi_v$ can be expressed in terms of $V_{lm}$
and $W_{lm}$ in an analogous way as $u$ and $v$ have been expressed
in the Equations A9 and A10.

In order to derive the time-evolution equations for $u$ and $v$
in spectral space we differentiate equations (A11-A14) with respect to 
time and obtain the following:

$${\partial b_r^{lm} \over \partial t} ={1 \over \sqrt{\alpha_{lm}}}
\left[\int_{-\frac{\pi}{2}}^{\frac{\pi}{2}}d\phi\,\cos\phi\, V_{lm}
\int_{0}^{2\pi}d\lambda\,\cos{m\lambda}\cdot\dot{v}
-\int_{-\frac{\pi}{2}}^{\frac{\pi}{2}} d\phi\,\cos\phi\,W_{lm}
\int_{0}^{2\pi}d\lambda\,\sin{m\lambda}\cdot\dot{u}\right],\quad\eqno(A23)$$

$${\partial b_i^{lm} \over \partial t} ={1 \over \sqrt{\alpha_{lm}}}
\left[\int_{-\frac{\pi}{2}}^{\frac{\pi}{2}}d\phi\,\cos{\phi}\, W_{lm}
\int_{0}^{2\pi}d\lambda\cos{m\lambda}\cdot\dot{u}
+\int_{-\frac{\pi}{2}}^{\frac{\pi}{2}} d\phi\,\cos{\phi}\,V_{lm}
\int_{0}^{2\pi}d\lambda\, \sin{m\lambda}\cdot\dot{v}\right],
\quad\eqno(A24)$$

$${\partial c_r^{lm} \over \partial t} = {1 \over \sqrt{\alpha_{lm}}}
\left[\int_{-\frac{\pi}{2}}^{\frac{\pi}{2}}d\phi\,\cos{\phi}\, V_{lm}
\int_{0}^{2\pi}d\lambda\,\cos{m\lambda}\cdot\dot{u}
+\int_{-\frac{\pi}{2}}^{\frac{\pi}{2}} d\phi\,\cos{\phi}\,W_{lm}
\int_{0}^{2\pi}d\lambda\sin{m\lambda} \cdot\dot{v}\right],
\quad\eqno(A25)$$

$${\partial c_i^{lm} \over \partial t} =-{1 \over \sqrt{\alpha_{lm}}}
\left[\int_{-\frac{\pi}{2}}^{\frac{\pi}{2}}d\phi\,\cos{\phi}\,W_{lm}
\int_{0}^{2\pi}d\lambda\,\cos{m\lambda}\cdot\dot{v}
+\int_{-\frac{\pi}{2}}^{\frac{\pi}{2}} d\phi\,\cos{\phi}\,V_{lm}
\int_{0}^{2\pi}d\lambda\sin{m\lambda} \cdot\dot{u}\right],
\quad\eqno(A26)$$

\noindent in which, $\dot{u}$ and $\dot{v}$ respectively denote the time 
derivatives of $u$ and $v$. Substituting $\dot{u}$ and $\dot{v}$ from 
equations (A21) and (A22) in the above equations (A23-A26), we obtain

$${\partial b_r^{lm} \over \partial t} = 
{1 \over \sqrt{\alpha_{lm}}}
\left\{\int_{-{\pi \over 2}}^{{\pi \over 2}}d\phi\,\cos{\phi} V_{lm}
\int_{0}^{2\pi}d\lambda\, \Xi_{v} \cos{m\lambda}-
\int_{-{\pi \over 2}}^{{\pi \over 2}} d\phi\,\cos{\phi}\,W_{lm}
\int_{0}^{2\pi}d\lambda\, \Xi_u \sin{m\lambda}\right\}
$$
%b_r^{lm}\left(\Xi\right)
$$
-\frac{1}{\sqrt{\alpha_{lm}}}\left[\int_{-\frac{\pi}{2}}^{\frac{\pi}{2}}
d\phi\,\cos{\phi}\, V_{lm}\int_{0}^{2\pi}d\lambda\,\cos{m\lambda}\cdot
\frac{\partial}{\partial\phi} \left(\frac{u^2+v^2}{2}+Gh\right)\right.$$
$$-\left.\int_{-\frac{\pi}{2}}^{\frac{\pi}{2}} d\phi\,\cos{\phi}\,W_{lm}
\int_{0}^{2\pi}d\lambda\,\sin{m\lambda}\cdot \frac{1}{\cos{\phi}}
\frac{\partial}{\partial\lambda} \left(\frac{u^2+v^2}{2}+Gh\right)\right],
$$
$$
= b_r^{lm}\left(\Xi\right)
-\frac{1}{\sqrt{\alpha_{lm}}}\left[\int_{-\frac{\pi}{2}}^{\frac{\pi}{2}}
d\phi\,\cos{\phi}\, V_{lm}\int_{0}^{2\pi}d\lambda\,\cos{m\lambda}\cdot
\frac{\partial}{\partial\phi} \left(\frac{u^2+v^2}{2}+Gh\right)\right.$$
$$-\left.\int_{-\frac{\pi}{2}}^{\frac{\pi}{2}} d\phi\,\cos{\phi}\,W_{lm}
\int_{0}^{2\pi}d\lambda\,\sin{m\lambda}\cdot \frac{1}{\cos{\phi}}
\frac{\partial}{\partial\lambda} \left(\frac{u^2+v^2}{2}+Gh\right)\right],
\quad\eqno(A27)$$

$${\partial b_i^{lm} \over \partial t} = 
%b_i^{lm}\left(\Xi\right)
{1 \over \sqrt{\alpha_{lm}}}
\left\{\int_{-{\pi \over 2}}^{{\pi \over 2}}d\phi\,\cos{\phi}\, W_{lm}
\int_{0}^{2\pi}d\lambda\, \Xi_u \cos{m\lambda}+
\int_{-{\pi \over 2}}^{{\pi \over 2}} d\phi\,\cos{\phi}\,V_{lm}
\int_{0}^{2\pi}d\lambda\, \Xi_v \sin{m\lambda}\right\}
$$
$$
- \frac{1}{\sqrt{\alpha_{lm}}}\left[\int_{-\frac{\pi}{2}}^{\frac{\pi}{2}}
d\phi\,\cos{\phi}\, W_{lm}\int_{0}^{2\pi}d\lambda\,\cos{m\lambda}\cdot
\frac{1}{\cos{\phi}} \frac{\partial}{\partial\lambda}
\left(\frac{u^2+v^2}{2} +Gh\right)\right.$$
$$+\left.\int_{-\frac{\pi}{2}}^{\frac{\pi}{2}} d\phi\,\cos{\phi}\,V_{lm}
\int_{0}^{2\pi}d\lambda\,\sin{m\lambda} \cdot
\frac{\partial}{\partial\phi} \left(\frac{u^2+v^2}{2}+Gh\right)\right],
$$
$$=
b_i^{lm}\left(\Xi\right)
- \frac{1}{\sqrt{\alpha_{lm}}}\left[\int_{-\frac{\pi}{2}}^{\frac{\pi}{2}}
d\phi\,\cos{\phi}\, W_{lm}\int_{0}^{2\pi}d\lambda\,\cos{m\lambda}\cdot
\frac{1}{\cos{\phi}} \frac{\partial}{\partial\lambda}
\left(\frac{u^2+v^2}{2} +Gh\right)\right.$$
$$+\left.\int_{-\frac{\pi}{2}}^{\frac{\pi}{2}} d\phi\,\cos{\phi}\,V_{lm}
\int_{0}^{2\pi}d\lambda\,\sin{m\lambda} \cdot
\frac{\partial}{\partial\phi} \left(\frac{u^2+v^2}{2}+Gh\right)\right],
\quad\eqno(A28)$$

$${\partial c_r^{lm} \over \partial t} = 
%c_r^{lm}\left(\Xi\right)
{1 \over \sqrt{\alpha_{lm}}}
\left\{\int_{-{\pi \over 2}}^{{\pi \over 2}}d\phi\,\cos{\phi}\, V_{nm}
\int_{0}^{2\pi}d\lambda\, \Xi_u \cos{m\lambda}+
\int_{-{\pi \over 2}}^{{\pi \over 2}} d\phi\,\cos{\phi}\,W_{lm}
\int_{0}^{2\pi}d\lambda\, \Xi_v \sin{m\lambda}\right\}
$$
$$
- \frac{1}{\sqrt{\alpha_{lm}}}\left[\int_{-\frac{\pi}{2}}^{\frac{\pi}{2}}
d\phi\,\cos{\phi}\, V_{lm}\int_{0}^{2\pi}d\lambda\,\cos{m\lambda}\cdot
\frac{1}{\cos{\phi}}\frac{\partial}{\partial\lambda}\left(\frac{u^2+v^2}{2}
+Gh\right)\right.$$
$$\left.+\int_{-\frac{\pi}{2}}^{\frac{\pi}{2}} d\phi\,\cos{\phi}\,W_{lm}
\int_{0}^{2\pi}d\lambda\,\sin{m\lambda}\cdot \frac{\partial}{\partial\phi}
\left(\frac{u^2+v^2}{2}+Gh\right)\right],$$
$$=c_r^{lm}\left(\Xi\right)
- \frac{1}{\sqrt{\alpha_{lm}}}\left[\int_{-\frac{\pi}{2}}^{\frac{\pi}{2}}
d\phi\,\cos{\phi}\, V_{lm}\int_{0}^{2\pi}d\lambda\,\cos{m\lambda}\cdot
\frac{1}{\cos{\phi}}\frac{\partial}{\partial\lambda}\left(\frac{u^2+v^2}{2}
+Gh\right)\right.$$
$$\left.+\int_{-\frac{\pi}{2}}^{\frac{\pi}{2}} d\phi\,\cos{\phi}\,W_{lm}
\int_{0}^{2\pi}d\lambda\,\sin{m\lambda}\cdot \frac{\partial}{\partial\phi}
\left(\frac{u^2+v^2}{2}+Gh\right)\right], \quad\eqno(A29)$$

$${\partial c_i^{lm} \over \partial t} = 
%c_i^{lm}\left(\Xi\right)
-{1 \over \sqrt{\alpha_{lm}}}
\left\{\int_{-{\pi \over 2}}^{{\pi \over 2}}d\phi\,\cos{\phi}\,W_{lm}
\int_{0}^{2\pi}d\lambda\, \Xi_v \cos{m\lambda}
+\int_{-{\pi \over 2}}^{{\pi \over 2}} d\phi\,\cos{\phi}\,V_{lm}
\int_{0}^{2\pi}d\lambda\, \Xi_u \sin{m\lambda}\right\}
$$
$$
- \frac{1}{\sqrt{\alpha_{lm}}}\left[\int_{-\frac{\pi}{2}}^{\frac{\pi}{2}}
d\phi\,\cos{\phi}\, W_{lm}\int_{0}^{2\pi}d\lambda\,\cos{m\lambda}\cdot
\frac{\partial}{\partial\phi} \left(\frac{u^2+v^2}{2}+Gh\right)\right.$$
$$\left.-\int_{-\frac{\pi}{2}}^{\frac{\pi}{2}} d\phi\,\cos{\phi}\,V_{lm}
\int_{0}^{2\pi}d\lambda\, \sin{m\lambda}\cdot
\frac{1}{\cos{\phi}} \frac{\partial}{\partial\lambda}\left(\frac{u^2+v^2}{2}
+Gh\right)\right]. $$
$$=c_i^{lm}\left(\Xi\right)
- \frac{1}{\sqrt{\alpha_{lm}}}\left[\int_{-\frac{\pi}{2}}^{\frac{\pi}{2}}
d\phi\,\cos{\phi}\, W_{lm}\int_{0}^{2\pi}d\lambda\,\cos{m\lambda}\cdot
\frac{\partial}{\partial\phi} \left(\frac{u^2+v^2}{2}+Gh\right)\right.$$
$$\left.-\int_{-\frac{\pi}{2}}^{\frac{\pi}{2}} d\phi\,\cos{\phi}\,V_{lm}
\int_{0}^{2\pi}d\lambda\, \sin{m\lambda}\cdot
\frac{1}{\cos{\phi}} \frac{\partial}{\partial\lambda}\left(\frac{u^2+v^2}{2}
+Gh\right)\right]. \quad\eqno(A30)$$

As in the case of the meteorological primitive equations, the shallow
water equations in the solar case admit high frequency gravity
waves as well as low frequency shearing flows. 
In order to increase the efficiency of the computation, we employ an
implicit scheme for the linear terms and an explicit scheme for the
nonlinear terms of the Equations (A27-A30). Therefore we sort out those 
terms and represent them in their corresponding spectral forms. 

Starting with the Equation (A27), we have the following linear term:
$$ -{1 \over \sqrt{\alpha_{lm}}}\left[ \int_{-\frac{\pi}{2}}^{\frac{\pi}{2}}
d\phi\cos{\phi} V_{lm}\int_{0}^{2\pi}d\lambda \cos{m\lambda} \left(G
{\partial h \over \partial\phi}\right)
-\int_{-\frac{\pi}{2}}^{\frac{\pi}{2}} d\phi\cos{\phi}W_{lm}
\int_{0}^{2\pi}d\lambda\sin{m\lambda}
\left(G{1 \over \cos{\phi}}{\partial h \over \partial\lambda}\right)\right].
$$
Substituting the expressions for ${1 \over \cos{\phi}}{\partial h 
\over \partial\lambda}$ and ${\partial h \over \partial\phi}$ from
Equations (A17) and (A18) into the above expression, and applying the
orthogonality relation from Equation (A7), we obtain
$$-{1 \over \sqrt{\alpha_{lm}}}\left[\int_{-\frac{\pi}{2}}^{\frac{\pi}{2}}
d\phi\cos{\phi} \, V_{lm}\int_{0}^{2\pi}d\lambda \cos{m\lambda} \sum_{{l_1}=0}^N
{\sum_{{m_1}=0}^{l_1}}^\star\sqrt{l_1\left({l_1}+1\right)}GV_{l_1m_1}
\left(a_r^{l_1m_1}\cos{m_1\lambda}+a_i^{l_1m_1}\sin{m_1\lambda}\right)\right.$$
$$ \left.-\int_{-\frac{\pi}{2}}^{\frac{\pi}{2}} d\phi\cos{\phi}W_{lm}
\int_{0}^{2\pi}d\lambda\sin{m\lambda} \sum_{l_1=0}^N 
{\sum_{m_1=0}^{l_1}}^\star\sqrt{\left(l_1+1\right)}GW_{l_1m_1}\cdot
\left(a_i^{l_1m_1}\cos{m_1\lambda}-a_r^{l_1m_1}\sin{m_1\lambda}\right)\right]
$$
$$ =-{\pi \over \sqrt{\alpha_{lm}}}\sum_{{l_1}=0}^N {\sum_{{m_1}=0}^{l_1}}^\star 
a_r^{l_1m_1} \sqrt{l_1\left({l_1}+1\right)} 
\int_{-\frac{\pi}{2}}^{\frac{\pi}{2}}d\phi\cos{\phi}\cdot G
\left(V_{lm}V_{l_1m_1} + W_{lm}W_{l_1m_1}\right) $$
$$=-a_r^{lm}G\sqrt{l\left(l+1\right)}. \hspace{4cm} \quad\eqno(A31)$$

A part of the nonlinear terms of equation (A27) contains the derivative
of $(u^2 + v^2 )/2$. Noting that $(u^2 + v^2 )/2$ is a scalar term which
is represented in spectral space as
$${u^2+v^2 \over 2}=  
\sum_{l=0}^N{\sum_{m=0}^l}^\star P_{lm}\left(d_r^{lm}\cos{m\lambda}+
d_i^{lm}\sin{m\lambda}\right), \quad\eqno(A32)$$
we can use similar formulae as given in expressions (A17) and (A18)
for the derivative of a scalar variable to evaluate the derivatives of
$(u^2 + v^2 )/2$. After a few steps of algebra, those terms can be
simplified to
$$
-\left[\int_{-\frac{\pi}{2}}^{\frac{\pi}{2}}
d\phi\,\cos{\phi}\, V_{lm}\int_{0}^{2\pi}d\lambda\,\cos{m\lambda}\cdot
\frac{\partial}{\partial\phi} \left(\frac{u^2+v^2}{2}\right)\right.$$
$$-\left.\int_{-\frac{\pi}{2}}^{\frac{\pi}{2}} d\phi\,\cos{\phi}\,W_{lm}
\int_{0}^{2\pi}d\lambda\,\sin{m\lambda}\cdot \frac{1}{\cos{\phi}}
\frac{\partial}{\partial\lambda} \left(\frac{u^2+v^2}{2}\right)\right] $$
$$ = -\left(G\cdot a_r^{lm}+d_r^{lm}\right)\sqrt{l\left(l+1\right)}.$$

Performing similar analysis for the nonlinear terms involving derivatives
of $(u^2 + v^2 )/2$ respectively in equations (A28-A30), we get
$$
- \left[\int_{-\frac{\pi}{2}}^{\frac{\pi}{2}}
d\phi\,\cos{\phi}\, W_{lm}\int_{0}^{2\pi}d\lambda\,\cos{m\lambda}\cdot
\frac{1}{\cos{\phi}} \frac{\partial}{\partial\lambda}
\left(\frac{u^2+v^2}{2} \right)\right.$$
$$+\left.\int_{-\frac{\pi}{2}}^{\frac{\pi}{2}} d\phi\,\cos{\phi}\,V_{lm}
\int_{0}^{2\pi}d\lambda\,\sin{m\lambda} \cdot
\frac{\partial}{\partial\phi} \left(\frac{u^2+v^2}{2}\right)\right] $$ 
$$=-\left(g\cdot a_i^{lm}+d_i^{lm}\right)\sqrt{l\left(l+1\right)}, $$

$$
- \left[\int_{-\frac{\pi}{2}}^{\frac{\pi}{2}}
d\phi\,\cos{\phi}\, V_{lm}\int_{0}^{2\pi}d\lambda\,\cos{m\lambda}\cdot
\frac{1}{\cos{\phi}}\frac{\partial}{\partial\lambda}\left(\frac{u^2+v^2}{2}
\right)\right.$$
$$\left.+\int_{-\frac{\pi}{2}}^{\frac{\pi}{2}} d\phi\,\cos{\phi}\,W_{lm}
\int_{0}^{2\pi}d\lambda\,\sin{m\lambda}\cdot \frac{\partial}{\partial\phi}
\left(\frac{u^2+v^2}{2}\right)\right]=0,$$

$$
- \left[\int_{-\frac{\pi}{2}}^{\frac{\pi}{2}}
d\phi\,\cos{\phi}\, W_{lm}\int_{0}^{2\pi}d\lambda\,\cos{m\lambda}\cdot
\frac{\partial}{\partial\phi} \left(\frac{u^2+v^2}{2}\right)\right.$$
$$\left.-\int_{-\frac{\pi}{2}}^{\frac{\pi}{2}} d\phi\,\cos{\phi}\,V_{lm}
\int_{0}^{2\pi}d\lambda\, \sin{m\lambda}\cdot
\frac{1}{\cos{\phi}} \frac{\partial}{\partial\lambda}\left(\frac{u^2+v^2}{2}
\right)\right]=0.$$

Recalling Equation (3) from \S2.1, we now write it as
$$ {\partial h \over \partial t} = \Psi -\delta, \quad\eqno(A33),$$
in which 
$$ \Psi = -h\delta-{u \over \cos\phi} {\partial h \over \partial\lambda} 
-v {\partial h \over \partial\phi}. \quad\eqno(A34)$$
To obtain the time-evolution equations for the spectral coefficients
involved in $h$, we differentiate Equations (A3) and (A4) with respect to 
time and using equation (A33) we obtain
$${\partial a_r^{lm} \over \partial t}=
{1 \over \sqrt{\alpha_{lm}}} \int_{-{\pi \over 2}}^{{\pi \over 2}}
d\phi P_{lm}\left(\phi\right)\cos{\phi}\int_{0}^{2\pi} d\lambda
\cos{m\lambda}\cdot\Psi $$
$$-{1 \over \sqrt{\alpha_{lm}}} \int_{-{\pi \over 2}}^{{\pi \over 2}}
d\phi P_{lm}\left(\phi\right)\cos{\phi}\int_{0}^{2\pi} d\lambda
\cos{m\lambda}\cdot\delta, \quad\eqno(A35)$$

$${\partial a_i^{lm} \over \partial t}= {1 \over \sqrt{\alpha_{lm}}} 
\int_{-{\pi \over 2}}^{\pi \over 2} d\phi P_{lm}\left(\phi\right)\cos{\phi}
\int_{0}^{2\pi} d\lambda\sin{m\lambda}\cdot\Psi\left(\phi,\lambda\right)
$$
$$
-{1 \over \sqrt{\alpha_{lm}}} \int_{-{\pi \over 2}}^{{\pi \over 2}}
d\phi P_{lm}\left(\phi\right)\cos{\phi}\int_{0}^{2\pi} d\lambda
\sin{m\lambda}\cdot\delta. \quad\eqno(A36)$$

Substituting in equations (A35) and (A36), the expression for $\delta$ 
from equation (A16), and using orthogonality of Legendre polynomials we 
get
$${\partial a_r^{lm} \over \partial t}=
{1 \over \sqrt{\alpha_{lm}}} \int_{-{\pi \over 2}}^{{\pi \over 2}}
d\phi P_{lm}\left(\phi\right)\cos{\phi}\int_{0}^{2\pi} d\lambda
\cos{m\lambda}\cdot\Psi +\sqrt{l\left(l+1\right)} b_r^{lm}, 
$$
$$=a_r^{lm}(\Psi) + \sqrt{l\left(l+1\right)}  b_r^{lm},
\quad\eqno(A37) $$  
$${\partial a_i^{lm} \over \partial t}= {1 \over \sqrt{\alpha_{lm}}} 
\int_{-{\pi \over 2}}^{\pi \over 2} d\phi P_{lm}\left(\phi\right)\cos{\phi}
\int_{0}^{2\pi} d\lambda\sin{m\lambda}\cdot\Psi\left(\phi,\lambda\right)
+\sqrt{l\left(l+1\right)}b_i^{lm}. 
$$
$$=a_i^{lm}(\Psi) + \sqrt{l\left(l+1\right)}  b_i^{lm},
\quad\eqno(A38)$$

The Equations (A27-A30) and (A37-A38) represent the full set of hydrodynamic 
shallow-water Equations (1-3) of \S2.1 in the spectral form. Like the
Equations (A37) and (A38), the Equations (A27-A30) can also be expressed
in compact form as

%$${\partial a_r^{lm} \over \partial t}=
%a_r^{lm}(\Psi) + \sqrt{l\left(l+1\right)}  b_r^{lm}, \quad\eqno(A39)$$

%$${\partial a_i^{lm} \over \partial t}=
%a_i^{lm}(\Psi) + \sqrt{l\left(l+1\right)}  b_i^{lm}, \quad\eqno(A40)$$

$${\partial b_r^{lm} \over \partial t} = 
%b_r^{lm}\left(\Xi\right)
%$$
%$$
%-\frac{1}{\sqrt{\alpha_{lm}}}\left[\int_{-\frac{\pi}{2}}^{\frac{\pi}{2}}
%d\phi\,\cos{\phi}\, V_{lm}\int_{0}^{2\pi}d\lambda\,\cos{m\lambda}\cdot
%\frac{\partial}{\partial\phi} \left(\frac{u^2+v^2}{2}+Gh\right)\right.$$
%$$-\left.\int_{-\frac{\pi}{2}}^{\frac{\pi}{2}} d\phi\,\cos{\phi}\,W_{lm}
%\int_{0}^{2\pi}d\lambda\,\sin{m\lambda}\cdot \frac{1}{\cos{\phi}}
%\frac{\partial}{\partial\lambda} \left(\frac{u^2+v^2}{2}+Gh\right)\right], $$
b_r^{lm}\left(\Xi\right) - \sqrt{l\left(l+1\right)}  d_r^{lm}
- G \sqrt{l\left(l+1\right)} a_r^{lm}, \quad\eqno(A39)$$

$${\partial b_i^{lm} \over \partial t} = 
%b_i^{lm}\left(\Xi\right)
%{1 \over \sqrt{\alpha_{lm}}}
%\left\{\int_{-{\pi \over 2}}^{{\pi \over 2}}d\phi\,\cos{\phi}\, W_{lm}
%\int_{0}^{2\pi}d\lambda\, \Xi_u \cos{m\lambda}+
%\int_{-{\pi \over 2}}^{{\pi \over 2}} d\phi\,\cos{\phi}\,V_{lm}
%\int_{0}^{2\pi}d\lambda\, \Xi_v \sin{m\lambda}\right\}
%$$
%$$
%- \frac{1}{\sqrt{\alpha_{lm}}}\left[\int_{-\frac{\pi}{2}}^{\frac{\pi}{2}}
%d\phi\,\cos{\phi}\, W_{lm}\int_{0}^{2\pi}d\lambda\,\cos{m\lambda}\cdot
%\frac{1}{\cos{\phi}} \frac{\partial}{\partial\lambda}
%\left(\frac{u^2+v^2}{2} +Gh\right)\right.$$
%$$+\left.\int_{-\frac{\pi}{2}}^{\frac{\pi}{2}} d\phi\,\cos{\phi}\,V_{lm}
%\int_{0}^{2\pi}d\lambda\,\sin{m\lambda} \cdot
%\frac{\partial}{\partial\phi} \left(\frac{u^2+v^2}{2}+Gh\right)\right], $$
b_i^{lm}\left(\Xi\right) - \sqrt{l\left(l+1\right)} d_i^{lm}
- G \sqrt{l\left(l+1\right)} a_i^{lm}, \quad\eqno(A40)$$

$${\partial c_r^{lm} \over \partial t} = 
%c_r^{lm}\left(\Xi\right)
%{1 \over \sqrt{\alpha_{lm}}}
%\left\{\int_{-{\pi \over 2}}^{{\pi \over 2}}d\phi\,\cos{\phi}\, V_{nm}
%\int_{0}^{2\pi}d\lambda\, \Xi_u \cos{m\lambda}+
%\int_{-{\pi \over 2}}^{{\pi \over 2}} d\phi\,\cos{\phi}\,W_{lm}
%\int_{0}^{2\pi}d\lambda\, \Xi_v \sin{m\lambda}\right\}
%$$
%$$
%- \frac{1}{\sqrt{\alpha_{lm}}}\left[\int_{-\frac{\pi}{2}}^{\frac{\pi}{2}}
%d\phi\,\cos{\phi}\, V_{lm}\int_{0}^{2\pi}d\lambda\,\cos{m\lambda}\cdot
%\frac{1}{\cos{\phi}}\frac{\partial}{\partial\lambda}\left(\frac{u^2+v^2}{2}
%+Gh\right)\right.$$
%$$\left.+\int_{-\frac{\pi}{2}}^{\frac{\pi}{2}} d\phi\,\cos{\phi}\,W_{lm}
%\int_{0}^{2\pi}d\lambda\,\sin{m\lambda}\cdot \frac{\partial}{\partial\phi}
%\left(\frac{u^2+v^2}{2}+Gh\right)\right],$$
c_r^{lm}\left(\Xi\right). \quad\eqno(A41)$$

Note that due to certain symmetry properties (antisymmetry here) of 
$V_{lm}$ and $W_{lm}$, 2nd and 3rd terms in the right hand side of
${\partial c_r^{lm} \over \partial t}$ equation (Equation (29)) get 
cancelled. Similar cancellation happens also in Equation (30) for
${\partial c_i^{lm} \over \partial t}$, which is reproduced in
compact form below, namely

$${\partial c_i^{lm} \over \partial t} = 
%c_i^{lm}\left(\Xi\right)
%-{1 \over \sqrt{\alpha_{lm}}}
%\left\{\int_{-{\pi \over 2}}^{{\pi \over 2}}d\phi\,\cos{\phi}\,W_{lm}
%\int_{0}^{2\pi}d\lambda\, \Xi_v \cos{m\lambda}
%+\int_{-{\pi \over 2}}^{{\pi \over 2}} d\phi\,\cos{\phi}\,V_{lm}
%\int_{0}^{2\pi}d\lambda\, \Xi_u \sin{m\lambda}\right\}
%$$
%$$
%- \frac{1}{\sqrt{\alpha_{lm}}}\left[\int_{-\frac{\pi}{2}}^{\frac{\pi}{2}}
%d\phi\,\cos{\phi}\, W_{lm}\int_{0}^{2\pi}d\lambda\,\cos{m\lambda}\cdot
%\frac{\partial}{\partial\phi} \left(\frac{u^2+v^2}{2}+Gh\right)\right.$$
%$$\left.-\int_{-\frac{\pi}{2}}^{\frac{\pi}{2}} d\phi\,\cos{\phi}\,V_{lm}
%\int_{0}^{2\pi}d\lambda\, \sin{m\lambda}\cdot
%\frac{1}{\cos{\phi}} \frac{\partial}{\partial\lambda}\left(\frac{u^2+v^2}{2}
%+Gh\right)\right]. $$
c_i^{lm}\left(\Xi\right). \quad\eqno(A42) $$

In order to evaluate the integrals efficiently and accurately, we use
Gaussian grids in both latitude and longitude directions. Denoting the 
non-uniformly spaced Gaussian grid points in $\phi$ by $\phi_j$ and the 
corresponding Gaussian weights by $w_j$, a $\phi$-integral can be 
expressed as 
$$I\left(\phi\right)=\int_{-{\pi \over 2}}^{{\pi \over 2}}
d\phi f\left(\phi\right)\approx\sum_j w_j f\left(\phi_j\right).
\quad\eqno(A43)$$
Similarly a $\lambda$-integral respectively involving a cosine and sine
functions can be expressed as:
$${I_1}\left(\lambda\right)=\int_{0}^{2\pi} d\lambda \,\cos m\lambda 
f\left(\lambda\right)\approx\sum_k w_k \cos m\lambda_k 
f\left( \lambda_k\right), \quad\eqno(A44)$$
$${I_2}\left(\lambda\right)=\int_{0}^{2\pi} d\lambda \,\sin m\lambda 
f\left(\lambda\right)\approx\sum_k w_k \sin m\lambda_k 
f\left( \lambda_k\right), \quad\eqno(A45)$$
in which $\lambda_k$ is the Gaussian grid in longitude and $w_k$ is
the corresponding weighting function. Using the expressions (A43-A45), the 
expansion coefficients for the height and the flow variables can be 
represented in Gaussian grids as:

$$a_r^{lm}={1 \over \sqrt{\alpha_{lm}}} \sum_j w_j P_{lm}
\left(\phi_j\right)\cos{\phi_j} \sum_{k=1}^{N_k} w_k \cos m\lambda_k
h\left(\phi_j,\lambda_k\right), \quad\eqno(A46)$$ 
$$a_i^{lm}={1 \over \sqrt{\alpha_{lm}}} \sum_j w_j P_{lm}
\left(\phi_j\right)\cos{\phi_j} \sum_{k=1}^{N_k} w_k \sin m\lambda_k
h\left(\phi_j,\lambda_k\right), \quad\eqno(A47)$$ 
$$b_r^{lm}=\sum_jw_j\cos{\phi_j}\left\{V_{lm}\left(\phi_j\right)
{\tilde{vc}}_m\left(\phi_j\right)-W_{lm}\left(\phi_j\right)
{\tilde{us}}_m\left(\phi_j\right)\right\}, \quad\eqno(A48)$$
$$b_i^{lm}=\sum_jw_j\cos{\phi_j}\left\{W_{lm}\left(\phi_j\right)
{\tilde{uc}}_m\left(\phi_j\right)+V_{lm}\left(\phi_j\right)
{\tilde{vs}}_m\left(\phi_j\right)\right\}i, \quad\eqno(A49)$$
$$c_r^{lm}=\sum_jw_j\cos{\phi_j}\left\{V_{lm}\left(\phi_j\right)
{\tilde{uc}}_m\left(\phi_j\right)+W_{lm}\left(\phi_j\right)
{\tilde{vs}}_m\left(\phi_j\right)\right\}, \quad\eqno(A50)$$
$$c_i^{lm}=\sum_jw_j\cos{\phi_j}\left\{-W_{lm}\left(\phi_j\right)
{\tilde{vc}}_m\left(\phi_j\right)+V_{lm}\left(\phi_j\right)
{\tilde{us}}_m\left(\phi_j\right)\right\}, \quad\eqno(A51)$$
in which
$${\tilde{uc}}_m\left(\phi_j\right) = \sum_{k=1}^{N_\lambda}
\cos(\lambda_k m) u\left(\phi_j,\lambda_k\right), $$
$${\tilde{us}}_m\left(\phi_j\right) = \sum_{k=1}^{N_\lambda}
\sin(\lambda_k m) u\left(\phi_j,\lambda_k\right), $$
$${\tilde{vc}}_m\left(\phi_j\right) = \sum_{k=1}^{N_\lambda}
\cos(\lambda_k m) v\left(\phi_j,\lambda_k\right), $$
$${\tilde{vs}}_m\left(\phi_j\right) = \sum_{k=1}^{N_\lambda}
\sin(\lambda_k m) v\left(\phi_j,\lambda_k\right). $$

\citet{swarzt96} solved the shallow-water equations in physical space
evaluating nonlinear terms in spectral space. We will be instead 
evolve the spectral coefficients following \citet{hack92},
so we construct a semi-implicit time evolution 
scheme, because the time step utilized by explicit time differencing
methods is limited by the speed of the fastest gravity wave. The semi-
implicit time integration removes this constraint by treating terms that
give rise to gravity waves implicitly, and the remaining terms
explicitly. 

Each of the Equations (A37-A42) has a linear term. 
We write all of these equations in centrally time differenced form and the
linear term averaged over time $\tau-1$ and $\tau+1$. We omit the
superscript of $lm$ from all the terms for focussing only on time 
evolution scheme.

$$
\frac{1}{2\Delta_\tau}\left[b_r^{\tau+1}-b_r^{\tau-1}\right] 
     = b_r^{\tau}(\Xi)-\sqrt{n(n+1)}d_r^\tau
     -G\sqrt{n(n+1)}\left(\frac{a_r^{\tau+1}+
     a_r^{\tau-1}}{2}\right), \quad\eqno(A52) $$

$$\frac{1}{2\Delta_\tau}\left[b_i^{\tau+1}-b_i^{\tau-1}\right] 
     = b_i^\tau(\Xi)-\sqrt{n(n+1)}d_i^\tau
     -G\sqrt{n(n+1)}\left(\frac{a_i^{\tau+1}+
     a_i^{\tau-1}}{2}\right), \quad\eqno(A53) $$

$$\frac{1}{2\Delta_\tau}\left[a_r^{\tau+1}-a_r^{\tau-1}\right] 
     = a_r^\tau(\Psi)+\sqrt{n(n+1)}\left(\frac{b_r^{\tau+1}+
     b_r^{\tau-1}}{2}\right), \quad\eqno(A54) $$

$$\frac{1}{2\Delta_\tau}\left[a_i^{\tau+1}-a_i^{\tau-1}\right] 
     =a_i^\tau(\Psi)+\sqrt{n(n+1)}\left(\frac{b_i^{\tau+1}+
     b_i^{\tau-1}}{2}\right). \quad\eqno(A55)$$

Equations (A52) and (A54) are coupled, similarly Equations (A53) and
(A55) are also coupled. We write Equations (A52) and (A54) in matrix 
form as

\renewcommand{\theequation}{A\arabic{equation}}
\begin{eqnarray}
\setcounter{equation}{56}
&&\left(\begin{array}{cc}g\Delta_\tau\sqrt{n\left(n+1\right)}&1\\
      1&-\Delta_\tau\sqrt{n\left(n+1\right)}\end{array}\right)
\left(\begin{array}{c}a_r^{\tau+1}\\
      b_r^{\tau+1}\end{array}\right) 
\\=&&
  \left(\begin{array}{c}
      2\Delta_{\tau}\left(b_r^\tau\left(\Xi\right)-\sqrt{n\left(n+1\right)}
      d_r^\tau\right)-g\Delta_\tau\sqrt{n\left(n+1\right)}a_r^{\tau-1}+
      b_r^{\tau-1}\\
      2\Delta_{\tau}a_r^\tau\left(\Psi\right)+
      \Delta_\tau\sqrt{n\left(n+1\right)}b_r^{\tau-1}+a_r^{\tau-1}
   \end{array}\right)
\end{eqnarray}
which can be written as
\begin{eqnarray}
\nonumber
a_r^{\tau+1}
&=&\frac{1}{\left[g\Delta_\tau^2n\left(n+1\right)+1\right]}
   \left[a_r^{\tau-1}\left\{-g\Delta_\tau^2n(n+1)+1\right\}\right.\\
   &&\left.+2\Delta_\tau\left\{a_r^\tau(\Psi)-\Delta_\tau n(n+1)d_r^\tau
   +\left(b_r^{\tau-1}+\Delta_\tau b_r^\tau(\Xi)\right)
   \sqrt{n(n+1)}\right\}\right]\\\nonumber
b_r^{\tau+1}
&=&\frac{1}{\left[g\Delta_\tau^2n\left(n+1\right)+1\right]}
   \left[b_r^{\tau-1}\left\{-g\Delta_\tau^2n(n+1)+1\right\}\right.
   \\&&\left.-2\Delta_\tau\left\{-b_r^\tau(\Xi)+\left(d_r^\tau
   +ga_r^{\tau-1}+g\Delta_\tau a_r^\tau(\Psi)\right)
   \sqrt{n(n+1)}\right\}\right]
\end{eqnarray}
Similarly, Equations (A53) and (A55) can be written as
\begin{eqnarray}
\nonumber
a_i^{\tau+1}
&=&\frac{1}{\left[g\Delta_\tau^2n\left(n+1\right)+1\right]}
   \left[a_i^{\tau-1}\left\{-g\Delta_\tau^2n(n+1)+1\right\}\right.\\
   &&\left.+2\Delta_\tau\left\{a_i^\tau(\Psi)-\Delta_\tau n(n+1)d_i^\tau
   +\left(b_i^{\tau-1}+\Delta_\tau b_i^\tau(\Xi)\right)
   \sqrt{n(n+1)}\right\}\right]\\\nonumber
b_i^{\tau+1}
&=&\frac{1}{\left[g\Delta_\tau^2n\left(n+1\right)+1\right]}
   \left[b_i^{\tau-1}\left\{-g\Delta_\tau^2n(n+1)+1\right\}\right.
   \\&&\left.-2\Delta_\tau\left\{-b_i^\tau(\Xi)+\left(d_i^\tau
   +ga_i^{\tau-1}+g\Delta_\tau a_i^\tau(\Psi)\right)
   \sqrt{n(n+1)}\right\}\right]
\end{eqnarray}
We discretize the Equations (A41) and (A42) in time for explicit 
evolution:
\begin{eqnarray}
c_r^{\tau+1}&=&c_r^\tau+\Delta_{\tau}c_r^{\tau-1}\left(\Xi\right)\\
c_i^{\tau+1}&=&c_i^\tau+\Delta_{\tau}c_i^{\tau-1}\left(\Xi\right)
\end{eqnarray}

We have presented above  all the equations we need to do time 
integration. At each time step we integrate equations (A58-63).
The details of the time evolution in each time-advancement involve 
sequentially the following steps:

\begin{enumerate}

\item From $b_r^{lm}$, $b_i^{lm}$, $c_r^{lm}$ and $c_i^{lm}$
      compute $u$ and $v$ in $(\theta,\phi)$ 
      Gaussian grid using Equations (A9) and (A10).

\item From $a_r^{lm}$ and $a_i^{lm}$ compute $h$ in
      $(\theta,\phi)$ Gaussian grid using Equation (A1).

\item From $b_r^{lm}$, $b_i^{lm}$, $c_r^{lm}$ and $c_i^{lm}$
      compute ${\bf \zeta}$ and ${\bf \delta}$ in 
      $(\theta,\phi)$ Gaussian grid using Equations (A15) and (A16).

\item From $u$, $v$, ${\bf \zeta}$ and ${\bf \delta}$ computed in
      steps 1 and 2, and $\omega_c$ given, compute 
      $\Xi_u$ and $\Xi_v$ using Equations (A19) and (A20).

\item Substitute $\Xi_u$ and $\Xi_v$, obtained from step 4, in Equations 
      (A11-A14) with $u$ and $v$ respectively to 
      evaluate $b_r^{lm}(\Xi)$, $b_i^{lm}(\Xi)$, $c_r^{lm}(\Xi)$ and 
      $c_i^{lm}(\Xi)$. 

\item Compute $\frac{1}{2}(u^2+v^2)$ from the already computed values
      of $u$ and $v$ from step 1, and then using sequentially the 
      Equations (A32), (A17), (A18), (A3) and (A4), compute $d_r^{lm}$ 
      and $d_i^{lm}$.

\item Compute the gradients of $\frac{1}{2}(u^2+v^2)$ in
      $(\theta,\phi)$ Gaussian grids using $d_r^{lm}$ and 
      $d_i^{lm}$ from step 6, using Equations (A17) and (A18).

\item Compute the gradients of $h$ in $(\theta,\phi)$ 
      Gaussian grids using $a_r^{lm}$ and $a_i^{lm}$ in 
      Equations (A17) and (A18).

\end{enumerate}

The above steps constitute the complete computations of right hand side of Equations 
(A37-A42). Using the right hand side we compute $b_r^{lm}$, $a_r^{lm}$, $b_i^{lm}$, 
$a_i^{lm}$ for the next time advancement from Equations (A58-A61), and $c_r^{lm}$ 
and $c_i^{lm}$ from Equations (A62) and (A63). We repeat the above eight steps 
forward as long as we want to continue the time marching.


\begin{thebibliography}{}
\bibitem[{{Arlt}, {Sule} \& {R\"udiger} (2005)}]{asr05}
{Arlt}, R., {Sule}, A. \& {R\"udiger}, G. 2005, \aap, 441, 1171

\bibitem[{{Cally} (2001)}] {cally01}
{Cally}, P. S. 2001, SolP, 199, 231

\bibitem[{{Cally}, {Dikpati} \& {Gilman} (2003)}] {cdg03}
{Cally}, P. S., {Dikpati}, M. \& {Gilman}, P. A. 2003, \apj, 582, 1190

\bibitem[{{Cally}, {Dikpati} \& {Gilman} (2004)}] {cdg04}
{Cally}, P. S., {Dikpati}, M. \& {Gilman}, P. A. 2004, ASP
Conf. Proc. 219, p.541, on ``Stars as suns : activity, 
evolution and planets'' IAU Symposium,  eds. A.K. Dupree and A.O. Benz. 
San Francisco, CA 

\bibitem[{{Chan}, {Liao} \& {Zhang} (2008)}] {clz2008}
{Chan}, K. H., {Liao}, X. \& {Zhang}, K. 2008, \apj, 682, 1392

\bibitem[{{Charbonneau}, {Dikpati} \& {Gilman} (1999)}] {cdg99}
{Charbonneau}, P., {Dikpati}, M. \& {Gilman}, P. A. 1999, \apj, 526, 523

\bibitem[{{Dikpati}, {Cally} \& {Gilman} (2004)}] {dcg2004}
{Dikpati}, M., {Cally}, P. S. \& {Gilman}, P. A. 2004, \apj, 610, 597

\bibitem[{{Dikpati} \& {Gilman} (1999)}] {dg99}
{Dikpati}, M. \& {Gilman}, P. A. 1999, \apj, 512, 417

\bibitem[{{Dikpati} \& {Gilman} (2001)}] {dg01}
{Dikpati}, M. \& {Gilman}, P. A. 2001, \apj, 551, 536

\bibitem[{{Dikpati} \& {Gilman} (2005)}] {dg05}
{Dikpati}, M. \& {Gilman}, P. A. 2005, \apj, 653, L193

\bibitem[{{Dikpati}, {Gilman} \& {Rempel} (2003)}] {dgr03}
{Dikpati}, M., {Gilman}, P. A. \& {Rempel}, M. 2003, \apj, 596, 680

\bibitem[{{Dziembowski} \& {Kosovichev} (1987)}] {dk87}
{Dziembowski}, W. \& {Kosovichev}, A. G. 1987, ACTA Astronomica, 37, 313

\bibitem[{{Fan}  {et~al.} (1999)}] {fzlf1999}
{Fan}, Y., {Zweibel}, E. G., {Linton}, M. G. \& {Fisher}, G. H. 1999, 
\apj, 521, 460

\bibitem[{{Garaud} (2001)}] {garaud01}
{Garaud}, P. 2001, MNRAS, 324, 68

\bibitem[{{Gelb} \& {Gleeson} (2001)}] {gelb01}
{Gelb}, A. \& {Gleeson}, J. P. 2001, Monthly Weather Rev., 129, 2346

\bibitem[{{Gilman} \& {Fox} (1997)}] {gf97}
{Gilman}, P. A. \& {Fox}, P. A. 1997, \apj, 484, 439

\bibitem[{{Gilman} \& {Dikpati} (2002)}] {gd02}
{Gilman}, P. A. \& {Dikpati}, M. 2002, \apj, 576, 1031

\bibitem[{{Gilman}, {Dikpati} \& {Miesch} (2007)}]{gdm07} {Gilman}, P.~A.,
{Dikpati}, M. \& {Miesch}, M.~S. 2007, \apjs, 170, 203

\bibitem[{{Hack} \& {Jakob} (1992)}] {hack92}
{Hack}, J. J. \& {Jakob}, R. 1992, NCAR Technical Note NCAT/TN-343+STR

\bibitem[{{Hollerbach} \& {Cally} (2009)}]{hc09}
{Hollerbach}, R. \& {Cally}, P.~S. 2009, Sol. Phys., 260, 251

\bibitem[{{Huang} \& {Pedlosky} (1998)}] {hp98}
{Huang}, R. X. \& {Pedlosky}, J. 1998, J. of Physical Oceanography, 29, 779

\bibitem[{{Hurlburt} \& {Hogan} (2008)}] {hh08}
{Hurlburt}, H. E. \& {Hogan}, P. J. 2008, Dynamics of Atmospheres and Oceans, 
45, 71

\bibitem[{{Miesch}, {Gilman} \& {Dikpati} (2007)}]{mgd07} {Miesch},
M.~S., {Gilman}, P.~A. \& {Dikpati}, M. 2007, \apjs, 168, 337

\bibitem[{{Pedlosky} (1987)}] {pedlosky87}
{Pedlosky}, J. 1987, Geophysical Fluid Dynamics (New York:Springer), 710

\bibitem[{{Philander} \& {Fedorov} (2003)}] {pf03}
{Philander}, S. G. \& {Fedorov}, A. 2003, Ann. Rev. of Earth and Planetary
Science, 31, 579

\bibitem[{{Poulin} \& {Flierl} (2003)}] {pf03b}
{Poulin}, S. G. \& {Flierl}, A. 2003, J. Phys. Oceanogr., 33, 2173

\bibitem[{{Rempel} \& {Dikpati} (2003)}] {rd03}
{Rempel}, M. \& {Dikpati}, M. 2003, \apj, 584, 524

\bibitem[{{Swarztrauber} (1996)}] {swarzt96}
{Swarztrauber}, P. N. 1996, Monthly Weather Rev., 124, 730

\bibitem[{{Watson} (1981)}] {watson81}
{Watson}, M. 1981, Geophys. Astrophys. Fluid Dyn., 16, 285

\bibitem[{{Zhang}, {Liao} \& {Schubert} (2003)}]{zls03} {Zhang},
K., {Liao}, X. \& {Schubert}, G. 2003, \apj, 585, 1124

\end{thebibliography}
\end{document}